\documentstyle[twocolumn,epsfig,fleqn,amstex,rotating]{mn}
\newcommand{\hi}{\mbox{H\ {\footnotesize I}}}
\newcommand{\hii}{\mbox{H\ {\footnotesize II}}}

\def\etal{{et al.\ }}

\def\lsim{~\rlap{\raise 0.4ex\hbox{$<$}}{\lower 0.7ex\hbox{$\sim$}}~}
\def\gsim{~\rlap{\raise 0.4ex\hbox{$>$}}{\lower 0.7ex\hbox{$\sim$}}~}
\def\dd{{\rm d}}

\def\bc{{\it \boldsymbol{c}}}
\def\bd{{\it \boldsymbol{d}}}
\def\bk{{\it \boldsymbol{k}}}

\def\bs{{\it \boldsymbol{s}}}
\def\bu{{\it \boldsymbol{u}}}
\def\bv{{\it \boldsymbol{v}}}
\def\bx{{\it \boldsymbol{x}}}
\def\by{{\it \boldsymbol{y}}}
\def\bz{{\it \boldsymbol{z}}}

\def\bI{{\bf I}}
\def\bN{{\bf N}}
\def\bS{{\bf S}}
\def\bW{{\bf W}}
\def\bSN{\bS+\bN}
\def\bNn{{\bf N}^{-1}}
\def\bSn{{\bf S}^{-1}}
\def\bSNn{\bS^{-1}+\bN^{-1}}

\def\omgl{\Omega_{\rm \Lambda}}
\def\omgb{\Omega_{\rm b}}
\def\omgm{\Omega_{\rm m}}
\def\fion{f_{\rm ion}}
\def\fv{f_{_{\rm V}}}
\def\nh{n_{_{\rm H}}}
\def\ne{n_{\rm e}}
\def\Te{T_{\rm e}}
\def\xe{x_{\rm e}}
\def\kb{k_{_{\rm B}}}
\def\tb{T_{\rm b}}
\def\tcmb{T_{_{\rm CMB}}}
\def\tk{T_{\rm k}}
\def\ts{T_{\rm s}}
\def\nuo{\nu_{\rm obs}}
\def\vion{V_{\rm ion}}

\def\lsun{{\rm L_\odot}}
\def\msun{{\rm M_\odot}}
\def\tvir{T_{\rm vir}}
\def\mvir{M_{\rm vir}}
\def\rvir{r_{\rm vir}}
\def\vvir{V_{\rm vir}}
\def\rl{\rho_{_{\rm l}}}
\def\rh{\rho_{_{\rm h}}}
\def\tcos{T_{\rm c}}
\def\tobs{T_{\rm o}}
\def\tres{T_{\rm r}}

\def\tru{T_{{\rm r}}^{u}}

\def\trv{T_{{\rm r}}^{v}}

\def\sgn{\sigma_{_{\rm N}}}
\def\Pk{{\mathcal P}_k}
\def\Pbk{{\mathcal P}_{\bk}}
\def\ni{{n_{_i}}}
\def\nl{{n_{_l}}}

\def\btcos{\mbox{\boldmath $\tcos$}}

\def\btobs{\mbox{\boldmath $\tobs$}}
\def\btobsli{\mbox{\boldmath $\tobs^{l,i}$}}

\def\btres{\mbox{\boldmath $\tres$}}
\def\btresli{\mbox{\boldmath $\tres^{l,i}$}}
\def\btru{\mbox{\boldmath $\tru$}}
\def\btrv{\mbox{\boldmath $\trv$}}

\def\btctru{\mbox{\boldmath $T_{\rm true}^l$}}
\def\btcrec{\mbox{\boldmath $T_{\rm rec}^l$}}

\begin{document}

\title[]
{De-contamination of cosmological 21-cm maps}
\author[Gleser {\it et al.}]{Liron Gleser$^{1}$, Adi Nusser$^{1}$, Andrew J. Benson$^{2}$\\\\
$^1$Physics Department, Technion, Haifa 32000, Israel\\
$^2$Theoretical Astrophysics, Caltech, MC130-33, 1200 E. California Blvd, Pasadena, CA 91125, U.S.A.}
\maketitle

\begin{abstract}
We present a method for extracting the expected cosmological 21-cm
signal from the epoch of reionization, taking into account
contaminating radiations and random instrumental noise. The method is
based on the maximum a-posteriori probability (MAP) formalism and
employs the coherence of the contaminating radiation along the
line-of-sight and the three-dimensional correlations of the
cosmological signal. We test the method using a detailed and
comprehensive modeling of the cosmological 21-cm signal and the
contaminating radiation. The signal is obtained using a high
resolution N-body simulation where the gas is assumed to trace the
dark matter and is reionized by stellar radiation computed from
semi-analytic galaxy formation recipes. We model contaminations to the
cosmological signal from synchrotron and free-free galactic
foregrounds and extragalactic sources including active galactic
nuclei, radio haloes and relics, synchrotron and free-free emission
from star forming galaxies, and free-free emission from dark matter
haloes and the intergalactic medium. We provide tests of the
reconstruction method for several rms values of instrumental noise
from $\sgn=1$ to $250$ mK. For low instrumental noise, the
recovered signal, along individual lines-of-sight, fits the true
cosmological signal with a mean rms difference of
$d_{\rm rms}\approx 1.7\pm 0.6$ for $\sgn=1$ mK, and
$d_{\rm rms}\approx 4.2\pm 0.4$ for $\sgn=5$ mK. The one-dimensional
power spectrum is nicely reconstructed for all values of $\sgn$
considered here, while the reconstruction of the two-dimensional power
spectrum and the Minkowski functionals is good only for noise levels
of the order of few mK.
\end{abstract}

\begin{keywords}
cosmology: theory --- diffuse radiation --- large-scale structure of
Universe --- intergalactic medium --- radio lines: general
\end{keywords}


\section {Introduction}
\label{sec:introduction}

The Wilkinson Microwave Anisotropy Probe (WMAP, Spergel \etal 2007)
polarization measurement of the cosmic microwave background (CMB)
indicates a value of $\tau=0.089\pm 0.030$ for the optical depth for
Thomson scattering with intervening free electrons in the
intergalactic medium (IGM). This value for the optical depth implies
the presence of ionizing radiation at redshifts $ z>10$ when the
Universe is less than half a Gyr old (see Meiksin 2007 for a
review). The ionizing radiation could be of varying nature depending
on the type of available sources as a function of time. X-ray
radiation emitted from mini black holes (Ricotti \& Ostriker 2004)
could be responsible for an early stage of (pre)-reionization. Because
of their large mean free path for photo-ionization, X-rays tend to
maintain the IGM at only partial reionization although full
reionization is possible near the sources (Zaroubi \etal 2007; Thomas
\& Zaroubi 2007). This possible X-ray pre-reionization is followed by
UV reionization which begins by the formation of individual \hii\
bubbles embedded the neutral medium. At a later stage \hii\ bubbles
overlap until complete reionization is obtained.

The CMB polarization measurement is proportional to the line-of-sight
integral of the ionized fraction and, therefore, is incapable of
constraining the details of the reionization process. Absorption
features in the spectra of high redshift quasars offer an alternative
probe of reionization. However, the dearth of quasars at the relevant
redshifts limits the applicability of this probe. Currently, a great
deal of effort is being made at observing the 21-cm signal from
neutral hydrogen from high redshifts. If the spin temperature of the
high redshift \hi\ is different from the CMB then redshifted 21-cm
radiation could be detected by radio observations at the relevant
frequency range (around 130 MHz for gas at $z\sim 10$). Maps of this
cosmological signal constrain the three-dimensional (3D) distribution
of \hi\ in the Universe, and, therefore, they could prove to be the
most useful probe of reionization. These maps will contain 3D
information on the distribution of \hi\ and the ionized fraction as a
function of time. Hopefully, in the near future radio telescopes like
the Low Frequency Array (LOFAR),\footnote{http://www.lofar.org} the
Precision Array to Probe Epoch of Reionization (PAPER),
\footnote{http://astro.berkeley.edu/$\sim$dbacker/EoR/} the Mileura
Widefield Array (MWA),
\footnote{http://www.haystack.mit.edu/ast/arrays/mwa/} the 21
Centimeter Array (21CMA, formerly
PaST),\footnote{http://cosmo.bao.ac.cn/project.html} and the Square
Kilometer Array (SKA), \footnote{http://www.skatelescope.org} will be
able to provide such 3D maps.

One of the main obstacles in extracting cosmological information from
redshifted 21 cm maps is the subtraction of non-cosmological
contaminations due to instrumental noise, ionospheric distortion,
galactic and extragalactic foregrounds. In the last few years an
effort was made to model the galactic foregrounds (e.g. Jelic \etal
2008; Santos, Cooray \& Knox 2005; Wang \etal 2006; Zaldarriaga,
Furlanetto \& Hernquist 2004), and the extragalactic foregrounds
(e.g. Cooray \& Furlanetto 2004; Di Matteo \etal 2002; Di Matteo,
Ciardi \& Miniati 2004; Jelic \etal 2008; Oh \& Mack 2003; Santos,
Cooray \& Knox 2005). In the first part of this work we model the
galactic and extragalactic foregrounds and produce 3D maps of the
extragalactic foregrounds.

The question of removing the foregrounds has been discussed
before. Most of the previous papers have focused on the angular power
spectrum of the signal and assumed rather limited spectral resolution
(e.g. Di Matteo \etal 2002; Di Matteo, Ciardi \& Miniati 2004; Oh \&
Mack 2003; Santos, Cooray \& Knox 2005; Zaldarriaga, Furlanetto \&
Hernquist 2004; Morales \& Hewitt 2004; Morales, Bowman \& Hewitt
2006). In general subtraction methods that have been suggested so far
includes three stages: bright source removal (e.g. Di Matteo \etal
2002; Di Matteo, Ciardi \& Miniati 2004), spectral fitting (Jelic
\etal 2008; Zaldarriaga, Furlanetto \& Hernquist 2004; Santos, Cooray
\& Knox 2005; Wang \etal 2006), and residual error subtraction
(Morales \& Hewitt 2004; Morales, Bowman \& Hewitt 2006). Using the
fact that the foregrounds which contaminate the 21-cm cosmological
signal have much smoother frequency spectra, Wang \etal 2006 developed
a method to remove the foregrounds along the line-of-sight. Based on
the maximum a-posteriori probability (MAP) formalism we develop a
method to reconstruct the 3D maps of the 21-cm cosmological signal
from the contaminated noisy data. This method also relies on the
smoothness of the contaminating radiation and further assumes a prior
for the correlation properties of the cosmological signal. On the
other hand, in this method bright sources does not get any special
treatment so no empty holes are left in the map. The random noise is
also included in the method so there is no need for the three stages
framework.

The paper is organized as follows. In \S~\ref{sec:signal}, we briefly
review the physics behind the 21-cm line. A brief description of the
simulation appears in \S~\ref{sec:simulation}. In
\S~\ref{sec:foregrounds} we present our models for the different
foregrounds, and in \S~\ref{sec:reconstruction} we describe the method
for the cosmological signal reconstruction. We conclude with a summary
and a discussion of the results in \S~\ref{sec:discussion}.


\section {The 21-cm cosmological signal}
\label{sec:signal}

\begin{figure*}
\centering
\resizebox{0.96\textwidth}{!}{\includegraphics{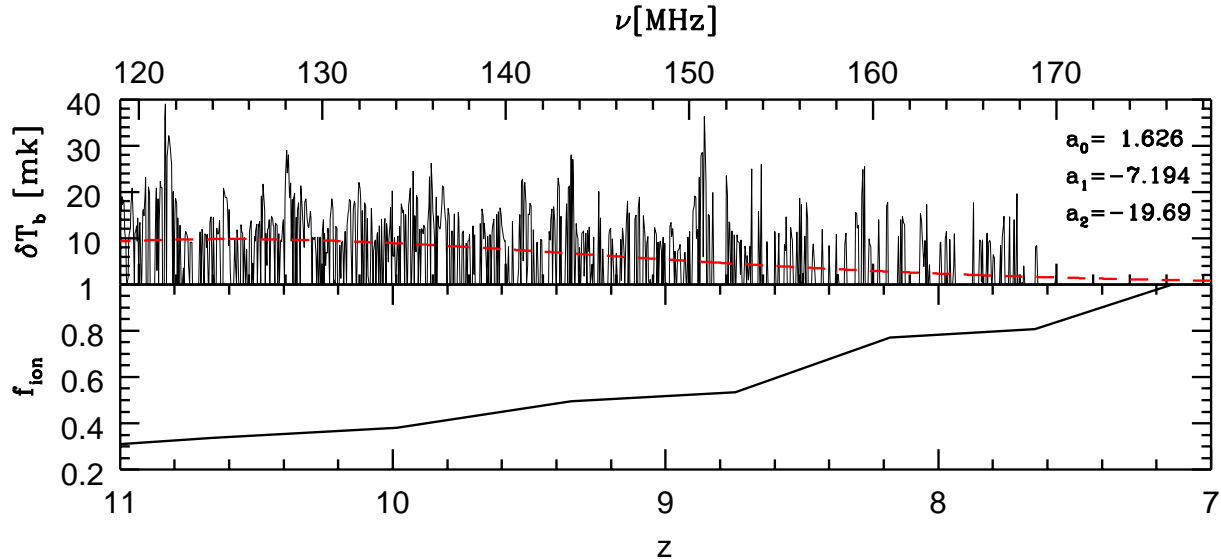}}
\caption{Top panel: the cosmological 21-cm signal along a typical
line-of-sight from $z=11$ ($\nu\approx 120$ MHz) to $z=7$ ($\nu\approx
180$ MHz). The dashed line shows the fit
$\delta\tb=\exp[a_{_0}+a_{_1}\ln(\nu/\nu_\ast)+a_{_2}\ln^2(\nu/\nu_\ast)]$,
where $\nu_\ast=150$ MHz. Bottom panel: The mean ionization fraction
in the simulation box as a function of redshift.}
\label{fig:dTb}
\end{figure*}

The 21-cm line is produced in the transition between the triplet and
singlet sub-levels of the hyperfine structure of the ground level of
neutral hydrogen atoms. This wavelength corresponds to a frequency of
1420 MHz and a temperature of $T_{*}=0.068$ K. The spin temperature,
$\ts$, is defined according to the relative population of the triplet,
$n_{_1}$, to the singlet, $n_{_0}$, sub-levels
\begin{equation}
\frac{n_{_1}}{n_{_0}}=3\exp\left(-\frac{T_{*}}{\ts}\right).
\label{eq:Ts}
\end{equation}
An \hi\ region would be visible against the CMB either in absorption
if $\ts<\tcmb$ or emission if $\ts>\tcmb$, where $\tcmb\approx
2.73(1+z)$ K is the CMB temperature. There are various mechanisms for
raising $\ts$ significantly above $\tcmb$ during the epoch of
reionization and hence a significant cosmological 21-cm signal is
expected.

Intensities, $I(\nu)$, at radio frequency are expressed in terms of
brightness temperature, defined as $\tb=I(\nu)c^2/(2\kb\nu^2)$, where
$c$ is the speed of light and $\kb$ is Boltzmann's constant. The
differential brightness temperature of the 21-cm line emission against
the CMB of a small patch of gas at redshift $z$ is (e.g. Ciardi \&
Madau 2003)
\begin{align}
\label{eq:dTb}
\delta\tb = & \ 16{\rm mK}\ x_{_{\rm H\ I}}(1+\delta)\left(1-\frac{\tcmb}{\ts}\right)\nonumber\\
 & \times\left(\frac{\omgb h}{0.02}\right)\left[\left(\frac{1+z}{10}\right)\left(\frac{0.3}{\omgm}\right)\right]^{1/2},
\end{align}
where $x_{_{\rm H\ I}}$ is the fraction of \hi\ in the patch, and
$\delta=\rho/\bar{\rho}-1$ is the density contrast of the gas. In this
paper we assume $\ts\gg\tcmb$ (e.g. Madau, Meiksin \& Rees 1997;
Nusser 2005a; Chuzhoy \& Zheng 2007). In the top panel of
Fig. \ref{fig:dTb} we present a typical cosmological 21-cm signal
along one line-of-sight from $z=11$ ($\nu\approx 120$ MHz) to $z=7$
($\nu\approx 180$ MHz) where the gas in the IGM is completely ionized
in our simulation (see \S~\ref{sec:simulation}). We also calculated
the mean differential brightness temperature,
$\left<\delta\tb\right>$, for this line-of-sight (dashed
line)\footnote{The fitting formula for the mean differential
brightness temperature is
$\delta\tb=\exp[a_{_0}+a_{_1}\ln(\nu/\nu_\ast)+a_{_2}\ln^2(\nu/\nu_\ast)]$,
where $\nu_\ast=150$ MHz. The same formula will be use in \S
\ref{sec:reconstruction} to fit the foregrounds contamination.}. In
the bottom panel we show the mean ionization fraction, $\fion$ of the
simulation box. As expected $\left<\delta\tb\right>$ decreases as
$\fion$ increases.


\section {The simulation}
\label{sec:simulation}

In order to produce 3D maps of the cosmological 21-cm signal and the
foregrounds, we use high resolution N-body simulation where the gas is
assumed to trace the dark matter and is reionized by stellar radiation
estimated from semi-analytic galaxy formation recipes (Benson \etal
2001 \& 2006). We calculate from the simulation the dark matter and
gas number density, the mass and number of dark matter haloes, and the
gas ionization fraction, all as a function of redshift.

The simulation box size is 141.3 h$^{-1}$Mpc comoving and it contains
$256^3$ particles, where each particle mass is $M_{\rm p}=2\times
10^{10}\ \msun$. We use a $\Lambda$CDM cosmology with $\omgm=0.3$,
$\omgl=0.7$, $h=0.7$, and $\sigma_8=0.9$. The galaxy formation model
includes Compton cooling, H$_2$ cooling and weak feedback from
supernova explosions, and the ionizing photons escape fraction from
galaxies is 0.15 (Benson \etal 2006). The reader is referred to that
paper, and references therein, for a full description of the
simulation.


\section{Foregrounds}
\label{sec:foregrounds}

The 21-cm cosmological signal will suffer from several sources of
contamination. Here we focus on the contamination which is produced by
foreground radiation emitted from galactic and extragalactic sources
and add a random instrumental noise with a rms value $\sgn$. We have
assumed that the necessary correction of ionospheric distortion, by
means of radio adaptive optics, and the calibration of the
time-variable gain, phase, and polarization of each antenna of the
radio telescope array, were already taking care of. Although the
contaminating radiation could be as much as $10^5$ times larger than
the cosmological signal (e.g. Gnedin \& Shaver 2004; McQuinn \etal
2006; Morales, Bowman \& Hewitt 2006), their typical power-law
dependence on frequency should allow us to disentangle the small scale
variations of the cosmological signal. In the following we describe
our modeling of the foreground contaminations.

Two physical processes are responsible for most contaminations:
synchrotron radiation and free-free emission. Synchrotron radiation
arises from acceleration of relativistic electrons in magnetic fields.
Assuming a power-law dependence of the electron number density on
energy, $N(E)\sim E^{-\gamma}$, where $E$ is the electron energy, one
obtains a synchrotron flux $I_{\rm syn}\sim\nu^{-\alpha}$ with
spectral index $\alpha=(\gamma-1)/2$. The corresponding brightness
temperature is $\tb\propto I_{\rm
syn}(\nu)\nu^{-2}\propto\nu^{-\beta}$, where $\beta=\alpha+2$. We will
consider synchrotron radiation from our Galaxy, star forming galaxies,
radio galaxies and radio haloes and relics in clusters. For each of
these systems we will assume a power-law synchrotron emissivity
consistent with the respective observational constraints (see below).

The emissivity of the free-free radiation from diffuse ionized gas is
(Rybicki \& Lightman 1979):
\begin{align}
\label{eq:emis}
\epsilon_\nu(\ne,\Te) = & \ 5.4\times 10^{-39}\ne^2\Te^{-1/2}g_{\rm ff}(\nu,\Te)\nonumber \\
 & \times e^{-h\nu/\kb\Te}{\rm\ erg\ cm^{-3}\ s^{-1}\ Hz^{-1}\ sr^{-1}},
\end{align}
where the Gaunt factor is approximately $g_{\rm
ff}(\nu,\Te)\approx11.96\Te^{0.15}\nu^{-0.1}$ in the radio regime
(Lang 1999), $\ne$ and $\Te$ are the electron number density and
temperature, and $\nu=\nuo(1+z)$ is the emitted frequency, with $\nuo$
the observed frequency. Therefore, the free-free flux density also has
a power-law dependence on frequency with spectral index $\alpha=0.1$
($\beta=2.1$).


\subsection{Galactic foregrounds}
\label{sec:galactic}

The galactic foregrounds are generated by synchrotron, free-free,
thermal and spinning dust emissions. In the relevant frequency range,
between $\sim$100 and $\sim$200 MHz, galactic synchrotron emission is
the most dominant foreground, responsible for $\sim$98.5\% of the
total galactic contamination, free-free emission from the diffuse
ionized hydrogen in the interstellar medium (ISM) contributes
$\sim$1.5\% (Shaver \etal 1999), and dust emission is negligible
(Platania \etal 2003; Reich, Testori \& Reich 2001). One should also
take into account radio recombination lines from our Galaxy. Since
these lines occur at specific frequencies, template spectra can be
used to remove them, without previous knowledge of their actual
intensity (e.g. Morales, Bowman \& Hewitt 2006).

The synchrotron emission process is probably due to radiation from
high energy cosmic-ray electrons above a few MeV in the galactic
magnetic field (e.g. Pacholczyk 1970; Banday \& Wolfendale 1990,
1991). There is less contamination in relatively smooth regions away
from the galactic plane and galactic loops. Following Wang \etal
(2006) we assume a running power-law in frequency for the galactic
synchrotron brightness temperature,
\begin{equation}
T_{\rm syn}=A_{\rm syn}\left(\frac{\nu}{\nu_\ast}\right)^{-\beta_{\rm syn}-\Delta\beta_{\rm syn}\log(\nu/\nu_\ast)},
\label{eq:Tsyn}
\end{equation}
where $A_{\rm syn}$ is synchrotron brightness temperature at
$\nu_\ast=150$ MHz, and $\beta_{\rm syn}$ and $\Delta\beta_{\rm syn}$
are the spectral index and spectral running index, respectively. Using
the 408 MHz all-sky continuum survey of Haslam \etal (1981, 1982),
Haverkorn, Katgert \& de Bruyn (2003) estimated the mean brightness
temperature at 408 MHz to be $\sim$33 K with temperature uncertainty
of $\sim$10\%. After subtraction of the $\sim$2.7 K contribution of
the CMB and the $\sim$3.1 K contribution of extragalactic sources
(e.g. Bridle 1967; Lawson \etal 1987; Reich \& Reich 1988), the
diffuse synchrotron galactic background is $\sim$27.2 K. Assuming a
spectral index of 2.74 (Platania \etal 2003) we estimate the
synchrotron brightness temperature at 150 MHz to be $A_{\rm
syn}=442.0\pm 44.2$ K. At high galactic latitudes, the brightness
temperature can drop to minimum of $\sim$200 K (Lawson \etal 1987;
Reich \& Reich 1988; Shaver \etal 1999). The estimations for the mean
spectral index around 150 MHz are ranged from 2.6 to 2.8 (e.g. Bridle
1967; Willis \etal 1977; Lawson \etal 1987; Reich \& Reich 1988;
Banday \& Wolfendale 1990, 1991; Tegmark \etal 2000; Platania \etal
2003) with indications for dispersion at each position on the sky, due
to distinct components along the line-of-sight (e.g. Lawson \etal
1987; Reich \& Reich 1988; Banday \& Wolfendale 1990, 1991; Shaver
\etal 1999). We choose spectral index of $\beta_{\rm syn}=2.7$ with
dispersion of 0.1 (Reich \& Reich 1988; Shaver \etal 1999), and
spectral running index of $\Delta\beta_{\rm syn}=0.1$ (Tegmark \etal
2000; Wang \etal 2006).

Free-free thermal emission appears in ionized regions in the ISM, with
electron temperature of $\Te>8000$ K. As for the galactic synchrotron,
we assume a running power-law in frequency for the galactic free-free
brightness temperature (Wang \etal 2006),
\begin{equation}
T_{\rm ff}=A_{\rm ff}\left(\frac{\nu}{\nu_\ast}\right)^{-\beta_{\rm ff}-\Delta\beta_{\rm ff}\log(\nu/\nu_\ast)},
\label{eq:Tff}
\end{equation}
where $A_{\rm ff}$ is the free-free brightness temperature at
$\nu_\ast=150$ MHz, and $\beta_{\rm ff}$ and $\Delta\beta_{\rm ff}$
are the spectral index and spectral running index,
respectively. Maintaining the 70:1 ratio between the synchrotron and
free-free emission (Shaver \etal 1999), we adopted $A_{\rm ff}=6.33\pm
0.63$ K assuming 10\% temperature uncertainty. At high frequencies
($\nu>10$ GHz) the brightness temperature spectral index is
$\beta_{\rm ff}=2.15$, while at low frequencies it drops to
$\beta_{\rm ff}=2.0$ due to optically thick self-absorption (Bennett
\etal 2003). Since the gas is optically thin above few MHz, the
brightness temperature spectrum between $\sim$100 and $\sim$200 MHz is
well determine with a spectral index of $\beta_{\rm ff}=2.10\pm 0.01$
(Shaver \etal 1999), and the spectral running index is
$\Delta\beta_{\rm ff}=0.01$ (Tegmark \etal 2000; Wang \etal 2006).


\subsection{Extragalactic foregrounds}
\label{sec:extragalactic}

The total emission of extragalactic foregrounds has been estimated
both directly and from integrated source counts. At 150 MHz, its
contribution to the contamination varies from $\sim$30 K (Willis \etal
1977; Cane 1979) to $\sim$50 K (Bridle 1967; Lawson \etal 1987; Reich
\& Reich 1988). These foregrounds produce $\sim$10\% of the total
contamination on average, but can reach $\sim$25\% at high galactic
latitudes, where the minimum brightness temperature of the diffuse
galactic emission drops to $\sim$200 K (Lawson \etal 1987; Reich \&
Reich 1988; Shaver \etal 1999).

Most foregrounds are due to radio point sources and relate to active
galactic nuclei (AGN) activity. Radio haloes and radio relics are also
significant foregrounds, but since they appear only in rich galaxy
clusters, there are rare. The remaining extragalactic foregrounds,
which are also the less significant ones, are synchrotron and
free-free emission from star forming galaxies, and free-free emission
from ionized hydrogen dark matter haloes and diffuse IGM.


\subsubsection{Radio emission from AGNs}
\label{sec:agn}

\begin{figure}
\centering
\mbox{\epsfig{figure=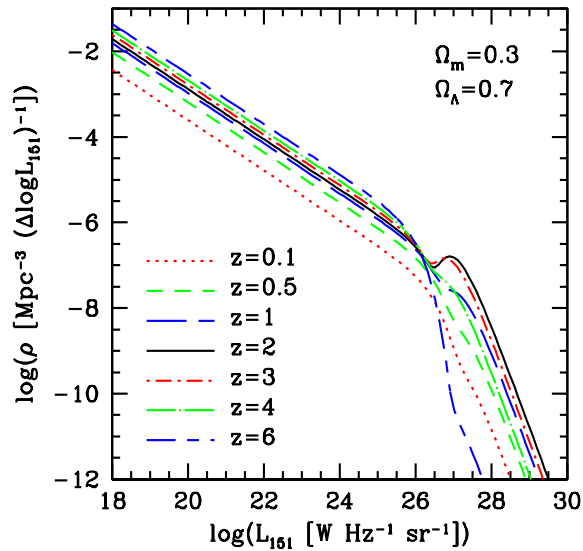,height=3.2in,width=3.2in}}
\caption{The Willott \etal (2001) RLF (model C) for $\Lambda$CDM
cosmology at seven redshifts from $z=0.1$ to $z=6$.}
\label{fig:wltgf}
\end{figure}

Radio emission from AGNs is linked to the accretion of gas on massive
black holes at the centers of galaxies. In order to incorporate this
contamination in the simulation we determine the number of AGNs in the
simulation box according to the radio luminosity function (RLF), and
assume that the AGN distribution follows the underlying mass density
field. The RLF is defined as the number of radio galaxies and quasars
per comoving volume per unit (base 10) logarithm of luminosity,
$\rho(L,z)$, where $L$ is the radio luminosity derived from the
flux-density $S$ and redshift per frequency and steradians.

We use the RLF model of Willott \etal (2001) to calculate the
background flux density from AGNs, assuming the luminosity of an AGN
has a power-law dependence on frequency with a spectral index of
$0.80\pm 0.15$. They used three redshift surveys of flux-limited
samples selected at the radio frequency range of our interest: 7CRS
(at 151 MHz with flux density $S_{151}\ge 0.5$ Jy), 3CRR (at 178 MHz
with $S_{178}\ge 10.9$ Jy), and 6CE (at 151 MHz with $2.0\ge
S_{151}<3.93$ Jy).

Willott \etal (2001) separated the radio sources to low- and
high-luminosity populations. A combination of the low-luminosity RLF,
$\rl$, and the high-luminosity RLF, $\rh$, gives the total RLF
\begin{equation}
\rho(L,z)=\rl+\rh.
\label{eq:rhottl}
\end{equation}
They checked three models for the redshift distribution. In our work
we use their redshift distribution which has a one-tailed Gaussian
rise to the peak redshift and then a one-tailed Gaussian decline at
higher redshifts (model C) with $\omgm=\omgl=0$ cosmology (for more
details see \S 3 in their paper).

The low-luminosity RLF is
\begin{equation}
\rl= \left\{
\begin{array}{ll}
\rho_{_{\rm l0}}\left(\frac{L}{L_{{\rm l}\ast}}\right)^{-\alpha_{\rm l}}\exp\left(\frac{-L}{L_{{\rm l}\ast}}\right)(1+z)^{k_{\rm l}} & z<z_{_{\rm l0}}\\
\rho_{_{\rm l0}}\left(\frac{L}{L_{{\rm l}\ast}}\right)^{-\alpha_{\rm l}}\exp\left(\frac{-L}{L_{{\rm l}\ast}}\right)(1+z_{_{\rm l0}})^{k_{\rm l}} & z\ge z_{_{\rm l0}}
\end{array}\right.,
\label{eq:rhol}
\end{equation}
where $\log(\rho_{_{\rm l0}})=-7.523$, $\log(L_{{\rm l}\ast})=26.48$,
$\alpha_{\rm l}=0.586$, $k_{\rm l}=3.48$, and $z_{_{\rm l0}}=0.710$.

The high-luminosity RLF is
\begin{equation}
\rh=\rho_{_{\rm h0}}\left(\frac{L}{L_{{\rm h}\ast}}\right)^{-\alpha_{\rm h}}\exp\left(-\frac{L_{{\rm h}\ast}}{L}\right)f_{_{\rm h}}(z),
\label{eq:rhoh}
\end{equation}
where $\log(\rho_{_{\rm h0}})=-6.757$, $\log(L_{{\rm h}\ast})=27.39$,
$\alpha_{\rm h}=2.42$, and the high-luminosity evolution function
$f_{_{\rm h}}(z)$ is
\begin{equation}
f_{_{\rm h}}(z)=\exp\left[-\frac{1}{2}\left(\frac{z-z_{_{\rm h0}}}{z_{_{\rm h1}}}\right)^2\right]
\label{eq:fh1}
\end{equation}
for $z<z_{_{\rm h0}}$, and
\begin{equation}
f_{_{\rm h}}(z)=\exp\left[-\frac{1}{2}\left(\frac{z-z_{_{\rm h0}}}{z_{_{\rm h2}}}\right)^2\right]
\label{eq:fh2}
\end{equation}
for $z\ge z_{_{\rm h0}}$, where $z_{_{\rm h0}}=2.03$, $z_{_{\rm
h1}}=0.568$, and $z_{_{\rm h2}}=0.956$.

To convert the RLF with $\omgm=\omgl=0$ cosmology, $\rho_{_0}$, to RLF
of $\Lambda$CDM cosmology, $\rho_{_{\rm \Lambda CDM}}$, we used the
following relation from Peacock (1985)
\begin{equation}
\rho_{_{\rm \Lambda CDM}}(L_{_{\rm \Lambda CDM}},z)\frac{\dd V_{_{\rm \Lambda CDM}}}{\dd z}=\rho_{_0}(L_{_0},z)\frac{\dd V_{_0}}{\dd z},
\label{eq:rhoconv}
\end{equation}
where $V_{_{\rm \Lambda CDM}}$ and $V_{_0}$ are the comoving volume in
$\Lambda$CDM and $\omgm=\omgl=0$ cosmologies, respectively. The
relation between the luminosities $L_{_{\rm \Lambda CDM}}$ and
$L_{_0}$ is $L_{_{\rm \Lambda CDM}}/L_{_0}=(D_{_{\rm \Lambda
CDM}}/D_{_0})^2$, where $D_{_{\rm \Lambda CDM}}$ and $D_{_0}$ are the
corresponding distances in the two cosmologies (see Peacock 1985;
Dunlop \& Peacock 1990 for more details). In Fig. \ref{fig:wltgf} we
present the RLF for $\Lambda$CDM cosmology at six redshifts, from 0.1
to 6.


\subsubsection{Radio haloes \& relics}
\label{sec:haloes}

\begin{figure}
\centering
\mbox{\epsfig{figure=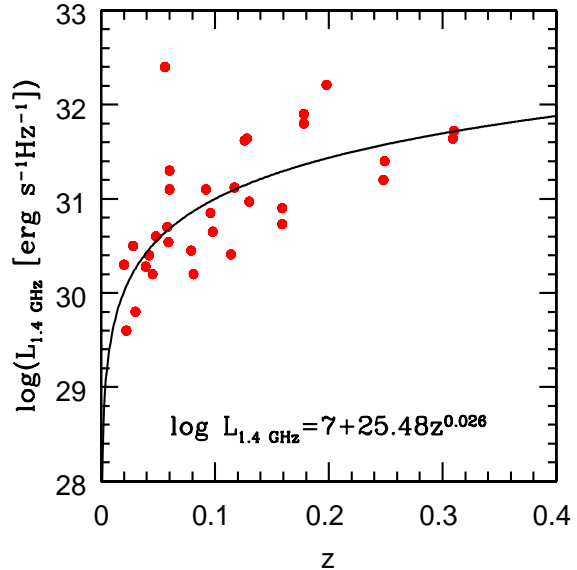,height=3.2in,width=3.2in}}
\caption{The relic radio luminosity at 1.4 GHz as a function of the
cluster redshift, where the full circles are the data from Giovannini
\& Feretti (2004), and the solid line is our model best fit.}
\label{fig:rlcgf}
\end{figure}

Diffuse non-thermal radio emission from the intra-cluster medium (ICM)
is observed in about 30 percent of rich galaxy clusters, which can be
found in dark matter haloes with $M_{\rm halo}\gsim 5\times
10^{14}\msun$. These sources are classified as {\it radio haloes} when
their morphology is regular, centered on and resembling the X-ray
emissivity, or {\it radio relics} when they are irregular, of
elongated shapes, and located at the periphery of the cluster
(e.g. Giovannini \& Feretti 2004). For weak magnetic fields ($B<0.5
\mu$G), the inverse Compton losses are dominant, while for stronger
magnetic fields, synchrotron emission is dominant (e.g. Cassano \etal
2006).

Cassano \etal (2006) studied the statistics of giant radio haloes
(GRHs) from electron re-acceleration models. We use this study to
determine the probability that a halo in the simulation hosts a GRH of
a given radio emissivity, as function of the halo mass and redshift
(see their Fig. 17a). The radio halo luminosity at 1.4 GHz is
\begin{equation}
L_{\rm 1.4\ GHz}=(4.85\pm 1.64)\times10^{30}\left(\frac{\mvir}{10^{15}\msun}\right)^{\alpha_{_{\rm M}}} {\rm erg\ s^{-1}Hz^{-1}},
\label{eq:L14grh}
\end{equation}
where $\mvir$ is the virial mass of the halo in $\msun$, and
$\alpha_{_{\rm M}}=2.9\pm 0.4$. The radio halo luminosity per
steradian at frequency $\nu$ is
\begin{equation}
L_{_{\rm GRH}}(\nu)=\frac{L_{\rm 1.4\ GHz}}{4\pi}\left(\frac{\nu}{{\rm 1.4\ GHz}}\right)^{-\alpha_\nu} {\rm erg\ Mpc^{-3}s^{-1}Hz^{-1}sr^{-1}},
\label{eq:emis_grh}
\end{equation}
where $\alpha_\nu=1.2$ (Cassano \etal 2006).

Radio relics are probably related to shock waves generated by cluster
mergers. These shock fronts can revive old radio ghosts when the
thermal pressure is much higher than the magnetic field pressure
(En{\ss}lin \& Goptal-Krishna 2001; Hoeft \etal 2004). Since the
available data and knowledge on radio relics are still poor, there is
no statistical model for the occurrence probability and radio
luminosity of relics as a function of the halo mass and
redshift. Therefore, we assume a 30\% probability for a radio relic to
exist in massive dark matter haloes, $M_{\rm halo}\gsim 5\times
10^{14}\msun$ (e.g. Di Matteo, Ciardi \& Miniati 2004). Giovannini \&
Feretti (2004) presented data from about 30 clusters of galaxies where
at least one radio relic is present. We use these data to roughly
estimate the relic radio luminosity at 1.4 GHz in erg
s$^{-1}$Hz$^{-1}$ as a function of the cluster redshift $z$ (see Fig
\ref{fig:rlcgf})
\begin{equation}
\log L_{\rm 1.4\ GHz}=7+(25.48\pm 0.27)z^{0.026\pm 0.004} {\rm erg\ s^{-1}Hz^{-1}}.
\label{eq:L14rlc}
\end{equation}
The radio relic luminosity per steradians at frequency $\nu$ is
\begin{equation}
L_{_{\rm RLC}}(\nu)=\frac{L_{\rm 1.4\ GHz}}{4\pi}\left(\frac{\nu}{{\rm 1.4\ GHz}}\right)^{-\alpha_\nu} {\rm erg\ s^{-1}Hz^{-1}sr^{-1}},
\label{eq:emis_rlc}
\end{equation}
where $\alpha_\nu=1.2$ (Kempner \etal 2004).


\subsubsection{Star forming galaxies}
\label{sec:glx emis}

\begin{table}
\caption{The relevant components of the ISM: the warm neutral medium
(WNM), the warm ionized medium (WIM), and the hot ionized medium
(HIM), neglecting the molecular medium (MM) and the cold neutral
medium (CNM) which are completely neutral and occupy less than 2\% of
the galaxy volume. $\Te$ is the gas temperature, $\nh$ is the hydrogen
number density, $\xe=\ne/\nh$ is the ionization fraction. $\fv$ is the
fraction of the interstellar volume, and $|Z|$ is the extent of the
gas component normal to the disk plane. The following values are based
on data from Heiles (2001) and T\"ullmann (2002).}
\vspace{1mm}
\begin{center}
\begin{tabular}{lccccc}
\hline
& $\Te$ [$10^4$ K] & $\nh$ [cm$^{-3}$] & $\xe$ & $\fv$ & $|Z|$ extent [kpc] \\
\hline
\hline
WNM & 0.05-0.8 & 0.1-0.5 & 0.2 & 0.4 & 0.5 \\
WIM & 0.8-1 & 0.05-0.08 & 1 & 0.1 & 5 \\
HIM & 10-1000 & 0.003-0.006 & 1 & 0.5 & 7 \\
\hline
\end{tabular}
\end{center}
\label{tbl:ism}
\end{table}

The gas in the ISM of star forming galaxies is mainly heated and
ionized by supernovae explosions, stellar winds and cosmic rays. We
calculated the free-free emission from the ISM components with a
significant ionized fraction: the warm neutral medium (WNM), the warm
ionized medium (WIM), and the hot ionized medium (HIM). While
neglecting the molecular medium (MM) and the cold neutral medium (CNM)
which are completely neutral and occupy less than 2\% of the
galaxy volume (e.g. Heiles 2001; T\"ullmann 2002).

In table \ref{tbl:ism} we present the physical properties of the
relevant ISM components for free-free emission. We used these data and
the free-free emissivity from equation (\ref{eq:emis}) to calculate
the luminosity of an ISM component $i$ of a single star forming galaxy
\begin{equation}
L_\nu^i(z)\approx\epsilon_\nu(\ne^i,\Te^i)\vion^i(z){\rm\ erg\ s^{-1}\ Hz^{-1}\ sr^{-1}},
\label{eq:lum_glx}
\end{equation}
where $\ne^i=\nh^i\xe^i$ is the free electrons number density of the
component, $\Te^i$ is the component temperature, and $\vion^i\approx
2\pi\fv r^2(z)Z(z)$ is the component ionized volume, where $r(z)$ is
the galaxy radius, $Z(z)\approx |Z|[r(z)/r_0]$ is the extent height of
the galaxy, and $r_0=15$ kpc is the approximate radius of the outer
stellar ring of the Milky Way (e.g. Helmi \etal 2003; Ibata \etal
2003; Yanny \etal 2003).

We adopt the Shen \etal (2003) galaxy radius distribution to draw random
radii in the simulation. Shen \etal (2003) used a complete set of
about 140000 galaxies from the Sloan Digital Sky Survey (SDSS) to
study the size distribution of galaxies as a function of
magnitude. They assume a log-normal distribution for the galaxy radius
\begin{equation}
f(r,\bar{r},\sigma_{_{\ln r}})=\frac{1}{\sqrt{2\pi}\sigma_{_{\ln r}}}\exp\left[-\frac{\ln^2\left(r/\bar{r}\right)}{2\sigma^2_{_{\ln r}}}\right]\frac{\dd r}{r},
\label{eq:f_radius}
\end{equation}
where the median, $\bar{r}(M)$, and the dispersion, $\sigma_{\ln
r}(M)$, are functions of the $r$-band Petrosian absolute magnitude,
$M$, in the range $-24\lsim M\lsim-16$. For late-type galaxies
\begin{equation}
\bar{r}(M)=\left[1+10^{-0.4(M-M_0)}\right]^{\beta-\alpha}10^{\gamma-0.4\alpha M}\ {\rm kpc},
\label{eq:r_median}
\end{equation}
and
\begin{equation}
\sigma_{\ln r}(M)=\sigma_2+\frac{\sigma_1-\sigma_2}{1+10^{-0.8(M-M_0)}},
\label{eq:r_sigma}
\end{equation}
where $\alpha=0.26$, $\beta=0.51$, $\gamma=-1.71$, $\sigma_1=0.45$,
$\sigma_2=0.27$, and $M_0=-20.91$ are fitting parameters. Following
Blanton \etal (2001, 2003), we approximate the galaxy luminosity
function in the $r$-band per unit magnitude by a Schechter function
\begin{align}
\label{eq:lum_fanc}
\Phi(M)\dd M = & \ 0.4\ln(10)\phi_\ast10^{-0.4(M-M_\ast)(\alpha+1)}\nonumber\\
 & \times\exp\left[-10^{-0.4(M-M_\ast)}\right]\dd M,
\end{align}
where at redshift $z=0.1$ the fitting parameters are
$\phi_\ast=0.0149\pm0.0004\ {\rm h^3 Mpc^{-3}}$,
$M_\ast-5\log_{10}{\rm h}=-20.44\pm0.01$, and
$\alpha=-1.05\pm0.01$. We scale the galaxy radii with redshift by
$r(z)\propto H(z)^{-1}$ (Ferguson \etal 2004).

In order to incorporate this contamination in the simulation we
determine the galaxy number counts for normal spiral galaxies using
the X-ray luminosity function from Ranalli \etal (2005), which based
on Takeuchi \etal (2003, 2004) far-infrared (FIR) luminosity function:
\begin{equation}
\varphi(L)=\varphi^\ast\left(\frac{L}{L^\ast}\right)^{1-\alpha}\exp\left[-\frac{1}{2\sigma^2}\log_{10}^2\left(1+\frac{L}{L^\ast}\right)\right]
\label{eq:LF}
\end{equation}
with $\varphi^\ast=0.0234\pm 0.0030 {\rm\ h^3Mpc^{-3}}$,
$\alpha=1.23\pm 0.04$, $\sigma=0.724\pm 0.01$, and $L^\ast=(4.4\pm
0.9)\times 10^8 {\rm\ h^{-2}}\lsun$. The galaxy number counts per
comoving volume at redshift $z$ is
\begin{equation}
N(z)=\int^{L_{\rm max}(z)}_{L_{\rm min}(z)}\dd\log L\varphi(\log L)\ {\rm h^3Mpc^{-3}}
\label{eq:Nz}
\end{equation}
where the minimum and maximum luminosities at $z=0$ are $L_{\rm
min}=10^{39}{\rm\ erg\ s^{-1}}$ and $L_{\rm max}=10^{43}{\rm\ erg\
s^{-1}}$, respectively. We consider luminosity evolution of $L\propto
(1+z)^{2.7}$ (Norman \etal 2004; Ranalli \etal 2005). We further
assume that the galaxy distribution follows the underlying mass
density field.


We use the calculation for free-free emission from star forming
galaxies to estimate their synchrotron emission. Shaver \etal (1999)
determine a $\sim$70:1 ratio between galactic synchrotron and
free-free emission at 150 MHz. Therefore, we assume that the
synchrotron emission is $70\pm 10$ times greater then the free-free
emission at 150 MHz. The spectral index of the synchrotron emission
depends on the relativistic electrons density distribution which
varies from galaxy to galaxy and within each galaxy. Hummel (1991)
analyzed the low frequency radio continuum data from 27 spiral
galaxies with deferent galaxy inclinations. He found that the mean
synchrotron spectral index of a galaxy at low frequencies ($\nu\le
700$ MHz) varies from $\sim$0.2 to $\sim$0.9 with a mean value of
$\alpha=0.56\pm 0.15$. Since our frequency range of interest is
limited to the vicinity of 150 MHz (100-200 MHz), we simplify the
frequency dependence by using a mean spectral index of $\alpha=0.6$
for all galaxies.


\subsubsection{Free-free emission from haloes and IGM}
\label{sec:ff emis}

The free-free emission from haloes comes from ionized gas which is
confined to dark matter haloes, at the virial
temperature. Photo-ionized diffuse gas in the IGM, typically, has a
mean temperature of $10^4$ K. Therefore, only haloes with
$\tvir\ge10^4$ K are able to confine gas to their gravitational
potential well. The virial temperature of a halo of mass $M_{\rm
halo}$ is given by
\begin{equation}
\tvir=3.6\times 10^5\left(\frac{V_{\rm c}}{100\ {\rm km\
s^{-1}}}\right)^2{\rm K}\; ,
\label{eq:Tvir}
\end{equation}
where the halo circular velocity is
\begin{equation}
V_{\rm c}=\left[100\Omega_{\rm m}(z)H^2(z)(GM_{\rm halo})^2\right]^{1/3}.
\label{eq:Vc}
\end{equation}

Neglecting temperature variations across the halo, according to
equation (\ref{eq:emis}), the total free-free emissivity of the halo
gas is proportional to the spatial integral of $\ne^2$. Assuming
further that the gas density follows that of the dark matter, the
total emissivity becomes proportional to the variance of mass density
fluctuations in the halo. To determine the density variance of a halo
of mass, $M_{\rm halo}$, at redshift ,$z$, we use the improved halo
density profile from Navarro \etal (2004), which is based on the
Navarro, Frenk \& White (1996, 1997, hereafter NFW) density profile
\begin{equation}
\frac{\rho(r)}{\rho_{_{-2}}}=\exp\left\{-\frac{2}{\alpha}\left[\left(\frac{r}{r_{_{-2}}}\right)^\alpha-1\right]\right\},
\label{eq:rho}
\end{equation}
where $r_{_{-2}}=\rvir/c$ is the characteristic radius in which the
NFW density profile is proportional to $r^{-2}$, $\rvir$ is the virial
radius of the halo, and $c$ is the concentration parameter. To
calculate the concentration for arbitrary virial mass and redshift, we
used the Eke \etal (2001) code available at J. F. Navarro's home
page\footnote{A link to the ENS subroutines for calculation of
concentrations for haloes of arbitrary mass and redshift can be found
at J. F. Navarro's home page:
http://www.astro.uvic.ca/$\sim$jfn/mywebpage/home.html}. The density
at $r_{_{-2}}$ is related to the NFW characteristic density,
$\rho_{\rm s}$, by $\rho_{_{-2}}=\rho_{\rm s}/4$, where
\begin{equation}
\rho_{\rm s}=\frac{200\rho_{_{\rm crit}}c^3}{3\left[\ln(1+c)-c/(1+c)\right]},
\label{eq:rho_s}
\end{equation}
where
\begin{equation}
\rho_{_{\rm crit}}(z)=\frac{3H_0^2}{8\pi G}\frac{\Omega_0}{\Omega(z)}(1+z)^3,
\label{eq:rho_crit}
\end{equation}
where $H_0$ is the current value of Hubble's constant, and the halo
mean density is $200\rho_{_{\rm crit}}$. Following Navarro \etal
(2004) we choose $\alpha=0.17$. The mass density variance is
\begin{equation}
\rho_{_{\rm var}}^2=\frac{4\pi}{\vvir}\rho_{_{-2}}^2\int_0^{\rvir}\exp\left\{-\frac{4}{\alpha}\left[\left(\frac{r}{r_{_{-2}}}\right)^\alpha-1\right]\right\}r^2\dd r,
\label{eq:rho_rms}
\end{equation}
where $\vvir=(4\pi/3)\rvir^3$ is the halo's virial volume. At redshift
$z=0$, $\rho_{_{\rm var}}^2/\rho_{_{\rm crit}}^2\approx 10^8$ for
$M_{\rm halo}=10^8\ \msun$, and $\rho_{_{\rm var}}^2/\rho_{_{\rm
crit}}^2\approx 0.5\times 10^8$ for $M_{\rm halo}=10^{12}\ \msun$.

We apply the friends-of-friends (FoF) algorithm to identify haloes in
the simulation box. Demanding haloes to have 10 particles or more,
gives a mass threshold of $M_{\rm halo}\ge 2\times 10^{11}\msun$ for
haloes in the simulation. Haloes with lower masses are introduced in
the simulation according to the Sheth \& Tormen (1999) mass function.
These haloes are placed at the positions of simulation particles not
belonging to haloes identified by the FoF algorithm. We calculate the
amount of radiation emanating from each cell in the simulation box by
collecting the radiation from all the haloes with virial temperatures,
$\tvir\ge10^4$ K.

The fraction of diffuse ionized gas in the IGM is taken to equal the
fraction of mass in haloes with $\tvir<10^4$ K. The free-free emission
from diffuse ionized gas in the IGM has a small contribution to the
overall contamination. The cumulative specific intensity for the
diffuse free-free emission from the IGM is
\begin{equation}
I_\nu = \int\dd\chi\frac{\epsilon_\nu(\ne,\Te)}{(1+z)^4}{\rm\ erg\ cm^{-2}\ s^{-1}\ Hz^{-1}},
\label{eq:Ispec}
\end{equation}
where $\dd\chi=ca^{-1}\dd t(z)$ is the differential conformal distance
and $\epsilon_\nu(\ne,\Te)$ is the free-free emissivity from equation
(\ref{eq:emis}).


\subsection{Results}
\label{sec:fg results}

\begin{figure}
\centering
\mbox{\epsfig{figure=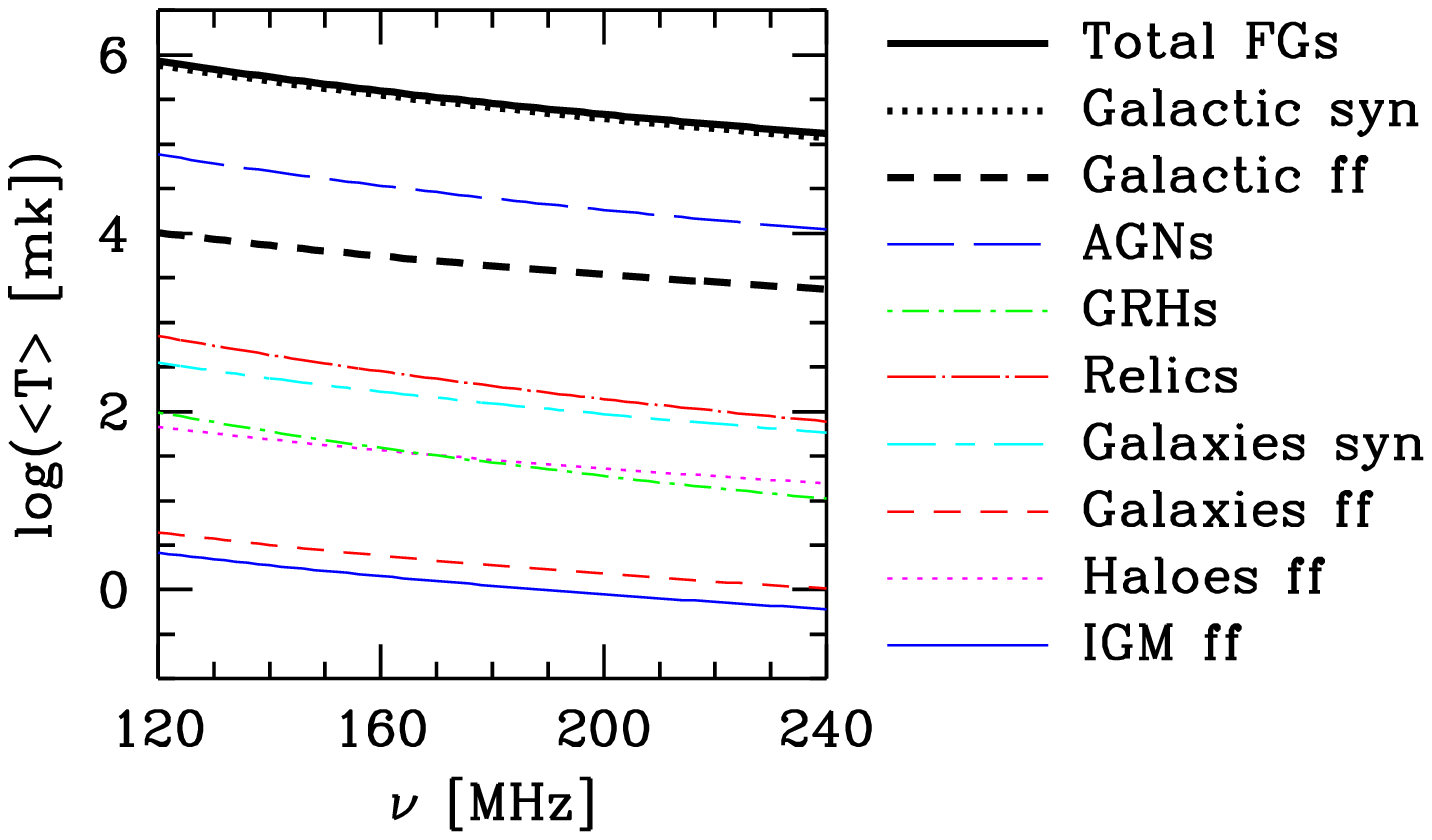,height=2.0in,width=3.2in}}
\caption{The mean brightness temperature of the different foregrounds
as a function of frequency. The total galactic and extragalactic
foregrounds (thick solid line), the galactic synchrotron emission
(thick dotted line), and the galactic free-free emission (thick dashed
line), the radio emission from AGNs (thin long-dashed line), radio
haloes (thin short-dash dotted line), and radio relics (thin long-dash
dotted line), the synchrotron (thin long-dashed short-dashed line) and
free-free emissions (thin short-dashed line) from star forming
galaxies, and the free-free emission from haloes (thin dotted line),
and the diffuse IGM (thin solid line).}
\label{fig:fgtmp}
\end{figure}

\begin{figure*}
\centering
\resizebox{0.96\textwidth}{!}{\includegraphics{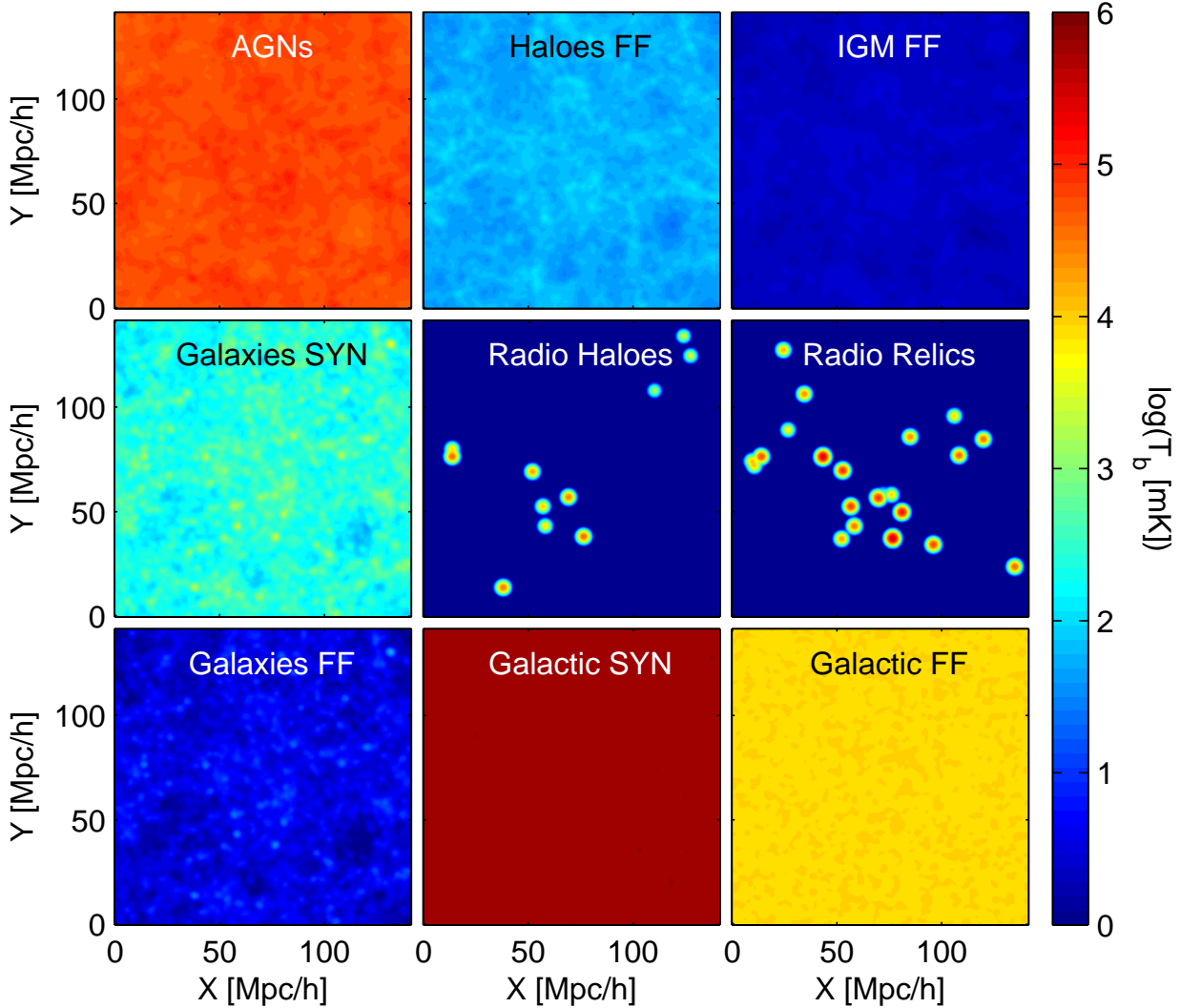}}
\caption{Maps of the brightness temperature foregrounds at
$\nu=129.78$ MHz: radio emission from AGNs (top-left), free-free from
haloes (top-middle), free-free from diffuse IGM (top-right),
synchrotron from star forming galaxies (center-left), emission from
radio haloes (center-middle) and radio relics (center-right),
free-free from star forming galaxies (bottom-left), galactic
synchrotron (bottom-middle), and galactic free-free (bottom-right).}
\label{fig:cexfg}
\end{figure*}

\begin{figure*}
\centering
\resizebox{0.96\textwidth}{!}{\includegraphics{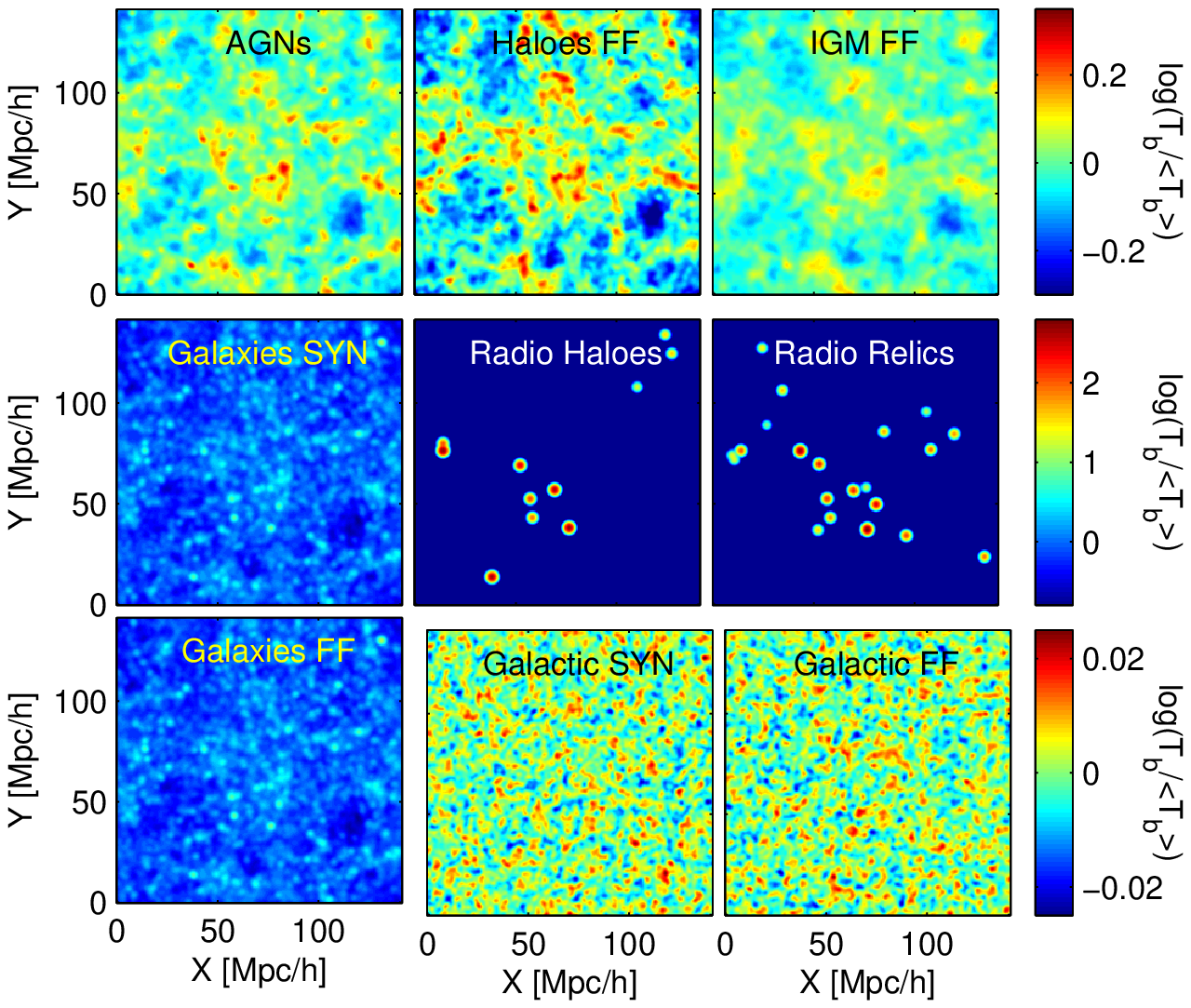}}
\caption{The same as Fig. \ref{fig:cexfg}, but for the brightness
temperature contrast in logarithmic scale,
$\log\left(\tb/\left<\tb\right>\right)$, where $\left<\tb\right>$ is
the mean brightness temperature in each panel. The three top panels
are scaled from -0.3 to 0.35. The three central panels and the
bottom-left panel are scaled from -0.8 to 2.8. The bottom-middle and
right panels are scaled from -0.025 to 0.025.}
\label{fig:cexfgd}
\end{figure*}

We have implemented the recipes for the foreground contaminations in
the simulation box. We work with 65536 lines-of-sight in a square
window of 141.3 h$^{-1}$Mpc (comoving) on the side and calculate the
foregrounds' total intensity in the frequency range $\nu=80-240$ MHz,
corresponding to the 21-cm signal at redshifts $z\approx 5-17$. In
Fig. \ref{fig:fgtmp} we show the mean brightness temperature of the
various foregrounds. As expected the most significant contamination is
the galactic synchrotron foreground ($\sim$89.5\%), next is the AGN
radio emission ($\sim 8.4-9.4$\%), and then the galactic free-free
($\sim 0.9-1.8$\%). The rest of the extragalactic foregrounds
contribute on average $\sim$0.15\%. We also tried several cases where
we changed the value of the galactic synchrotron emission from a
fairly low galactic foreground, which resemble high latitude regions
(Shaver \etal 1999), where the contribution of the galactic
synchrotron emission drops to $\sim$77\% of the total contamination
($A_{\rm syn}=150\pm 15$ K) and the galactic free-free emission adds
$\sim$1\% ($A_{\rm ff}=2.250\pm 0.225$ K), to an extremely high
galactic foreground, which might be found in radio loops (Reich \&
Reich 1988), where the galactic synchrotron emission contributes
$\sim$98\% of the total contamination ($A_{\rm syn}=2000\pm 200$ K)
and the galactic free-free adds $\sim$0.4\% ($A_{\rm ff}=6.33\pm 0.63$
K).

In Fig. \ref{fig:cexfg} we present logarithmic maps of the brightness
temperature of the integrated extragalactic foregrounds along the
lines-of-sight which contaminate the cosmological 21-cm signal at
$z=10$ ($\nu=129.78$ MHz). In Fig. \ref{fig:cexfgd} we present the
same logarithmic maps but for the brightness temperature contrast,
$\log\left(\tb/\left<\tb\right>\right)$, where $\left<\tb\right>$ is
the mean brightness temperature in each panel. As expected the most
dominant foreground is the galactic synchrotron (bottom-middle) and
next is the radio emission from AGN (top-left). The free-free
emissions from the IGM (top-right), haloes (top-middle) and star
forming galaxies (bottom-left) contribute very little to the
contamination. The emission from radio haloes (center-middle) and
relics (center-right), which seems to be insignificant on average, can
actually contribute significant contamination in individual
lines-of-sight. Since there are only a few haloes with $M_{\rm
halo}\gsim 5\times 10^{14}\msun$ in the simulation box, for most
lines-of-sight the contamination is practically zero, while in a few
lines-of-sight the brightness temperature contamination from radio
halo or relic can be as high as few $10^5$ mK.


\section{Signal reconstruction}
\label{sec:reconstruction}

The MAP formalism have been used to reconstruct the large-scale
structure of the Universe (e.g. Rybicki \& Press 1992; Fisher \etal
1995; Zaroubi \etal 1995). Based on this formalism, we develop a
method to reconstruct the 21-cm cosmological signal from the
contaminated data. The method relies on the smoothness of the
contaminating radiation along the frequency axis and an assumed prior
for the correlation properties of the cosmological signal. According
to Bayes theorem the conditional probability distribution function for
the occurrence of a 3D map, $\btcos$, of a 21-cm cosmological signal
given an observed data $\btobs$ is
\begin{align}
\label{eq:pl}
P(\btcos|\btobs) = & \ \frac{P(\btobs|\btcos)P(\btcos)}{P(\btobs)}\nonumber\\
 \propto & \ \exp\left\{-\frac{1}{2}\sum_{\bu}\sum_{\bv}\left[\btru[\bSN]^{-1}_{\bu,\bv}\btrv+\right.\right.\\
 & \!\!\left.\left.\left(\btcos-\bW\btres\right)^{\bu}[\bSNn]_{\bu,\bv}\left(\btcos-\bW\btres\right)^{\bv}\right]\right\},\nonumber
\end{align}
where $\bu$ and $\bv$ are position vector indices, and
$\bW\equiv\bS[\bSN]^{-1}$ is the Wiener filter. We assume that the
signal is approximately a Gaussian random field with an
auto-correlation function $\bS$ and that the (uncorrelated) noise is
Gaussian with a diagonal correlation function $\bN$ which is
$\bN_{\bu,\bv}=0$ for $\bu\neq\bv$ and $\bN_{\bu,\bu}=\sgn^2$ for
$\bu=\bv$ where $\sgn$ is the noise rms value assumed to be constant
(for more details see Appendix \ref{apx:probability}). We determine
$\btres$ as the residue of the observed 3D signal, $\btobs$, after
subtracting the approximated foregrounds. The $i$th element along the
line-of-sight $l$ of $\btres$ is
\begin{equation}
\btresli=\btobsli-\exp\left\{\sum_{m=0}^{n_{_m}}b_{_{l,m}}\left[\ln\left(\frac{\nu_{_i}}{\nu_{_\ast}}\right)\right]^m\right\},
\label{eq:T0}
\end{equation}
where we used a polynomial fitting in $\ln(\nu_{_i}/\nu_{_\ast})$ of
order $n_{_m}=2$ for the foregrounds, and $(\nu_{_i}/\nu_{_\ast})$ is
the ratio between the frequency in the $i$th bin and $\nu_{_\ast}=150$
MHz. The coefficients of the polynomial, $b_{_{l,m}}$, are determined
by minimization of the first term of the exponent in the rhs of the
conditional probability distribution function, equation (\ref{eq:pl}),
either in real space
\begin{equation}
q=\frac{1}{2}\sum_{\bu}\sum_{\bv}\btru[\bSN]^{-1}_{\bu,\bv}\btrv
\label{eq:fl}
\end{equation}
or in $k$-space
\begin{equation}
{\mathcal Q}=\frac{1}{2}\sum_\bk\frac{{\mathcal T}^2_{{\boldsymbol{\rm r}},\bk}}{\Pbk+\sgn^2},
\label{eq:Fl}
\end{equation}
where ${\mathcal T}_{{\boldsymbol{\rm r}},\bk}$ is the $\bk$ vector
element of the 3D Fourier transform of $\btres$, and $\Pbk$ is the 3D
power spectrum of the brightness temperature field. To find the best
set of $b_{_{l,m}}$ coefficients, we minimize $\mathcal Q$ by an
iterative steepest descent algorithm using the Newton direction as the
descent direction
\begin{equation}
d=-\left(\nabla^2{\mathcal Q}\right)^{-1}\nabla{\mathcal Q},
\label{eq:d_decsent}
\end{equation}
where $\nabla{\mathcal Q}$ and $\nabla^2{\mathcal Q}$ are the gradient
and the Hessian of $\mathcal Q$, respectively (for more details see
Appendix \ref{apx:contamination}). Then by solving the equation
$\partial P/\partial\btcos=0$, one gets the ``optimal" signal
\begin{equation}
\btcos=\bW\btres=\bS[\bSN]^{-1}\btres,
\label{eq:Tcos}
\end{equation}
or in $k$-space, $\partial P/\partial{\mathcal T}_{{\boldsymbol{\rm
c}},\bk}=0$,
\begin{equation}
{\mathcal T}_{{\boldsymbol{\rm c}},\bk}=\left(\frac{\Pk}{\Pk+\sgn^2}\right)^\eta{\mathcal T}_{{\boldsymbol{\rm r}},\bk},
\label{eq:Tcosk}
\end{equation}
where ${\mathcal T}_{{\boldsymbol{\rm c}},\bk}$ is the $\bk$ element
of the 3D Fourier transform of $\btcos$, $\Pk=\left<\Pbk\right>_k$ is
the one-dimensional (1D) mean power spectrum, where $k=|\bk|$, and the
Wiener filter parameter $\eta$ determines the strength of the Wiener
filter. According to the Wiener filter formalism $\eta=1$, but in this
case one does not recover all the small scale power since
\begin{align}
\label{eq:Pk rec}
\Pk^{\rm rec} & =\left<\left({\mathcal T}_{{\boldsymbol{\rm c}},\bk}\right)^2\right>=\left(\frac{\Pk}{\Pk+\sgn^2}\right)^2\left<\left({\mathcal T}_{{\boldsymbol{\rm r}},\bk}\right)^2\right>\nonumber\\
 & \approx\left(\frac{\Pk}{\Pk+\sgn^2}\right)^2\left(\Pk+\sgn^2\right)=\left(\frac{\Pk}{\Pk+\sgn^2}\right)\Pk.
\end{align}
Using $\eta=0.5$ eliminates the power suppression at small scales and
$\Pk^{\rm rec}\approx\Pk$, but slightly spoils the reconstruction
results for each individual line-of-sight.


\subsection{Results}
\label{sec:rec results}

\begin{figure}
\centering
\mbox{\epsfig{figure=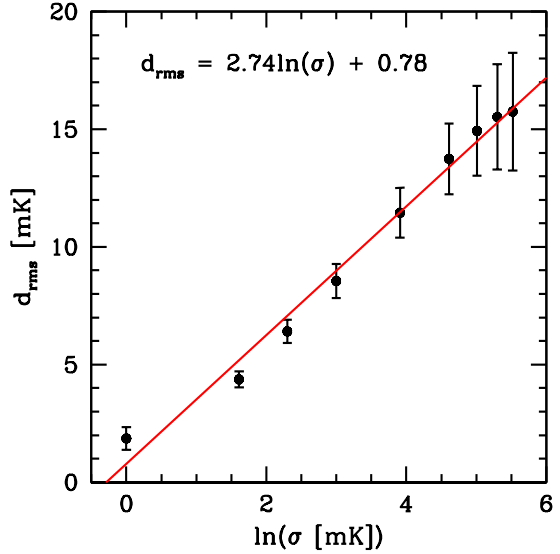,height=3.2in,width=3.2in}}
\caption{The mean $d_{\rm rms}$ for all the lines-of-sight in redshift
space as a function of the instrumental noise. The full circles are
for nine values of instrumental noise, $\sgn=1$, $5$, $10$, $20$,
$50$, $100$, $150$, $200$, and $250$ mK, with $\eta=1$. The solid line
is the best linear fit for these data points in logarithmic scale,
$\left<d_{\rm rms}\right>=2.74\ln\sgn+0.78$.}
\label{fig:dsig}
\end{figure}

\begin{figure*}
\centering
\resizebox{0.96\textwidth}{!}{\includegraphics{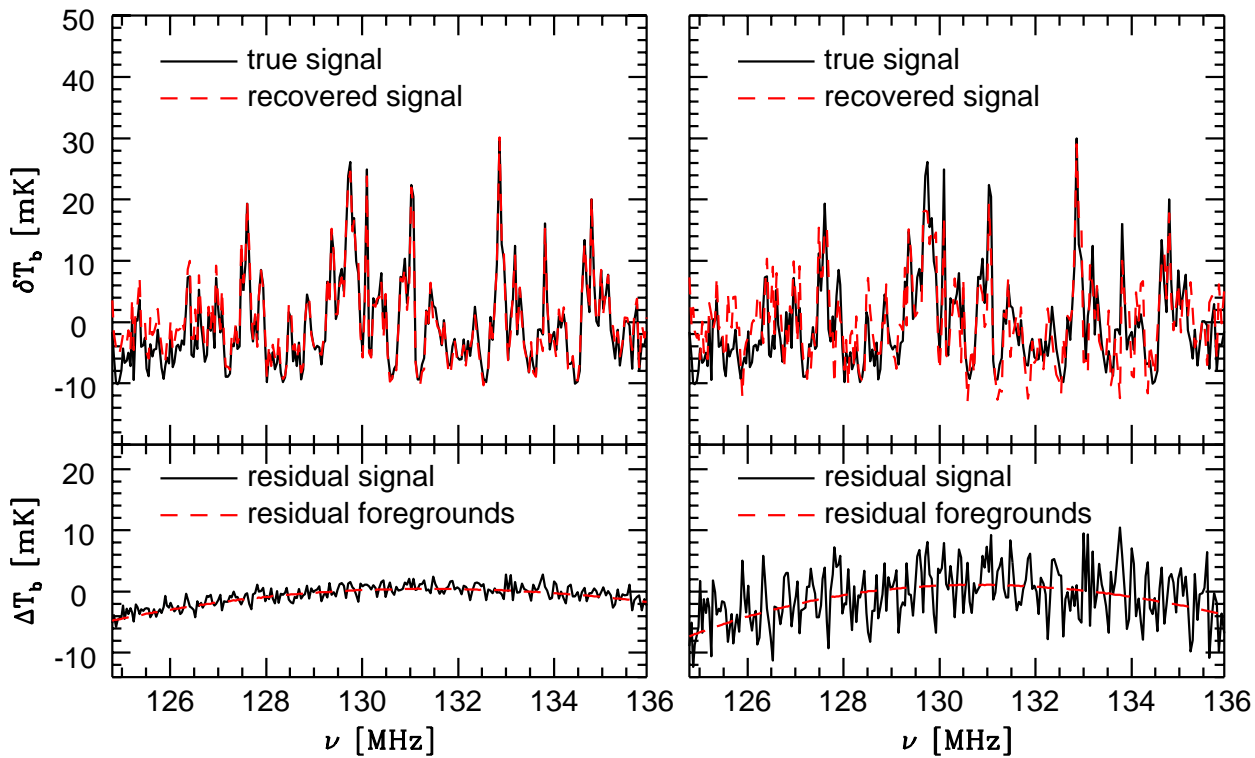}}
\caption{A single line-of-sight signal reconstruction in redshift
space for $\sgn=1$ mK and $d_{\rm rms}=1.77$ (left panels) and
$\sgn=5$ mK and $d_{\rm rms}=4.50$ (right panels). Top panels: the
brightness temperature, $\delta\tb$, of the true cosmological signal
(solid line) and the recovered signal (dashed line). Bottom panels:
the residue, $\Delta\tb$, between the true and recovered signals
(solid line) and between the true and the fitted foregrounds (dashed
line). For all cases $\eta=0.5$.}
\label{fig:tc_avg}
\end{figure*}

\begin{figure*}
\centering
\resizebox{0.96\textwidth}{!}{\includegraphics{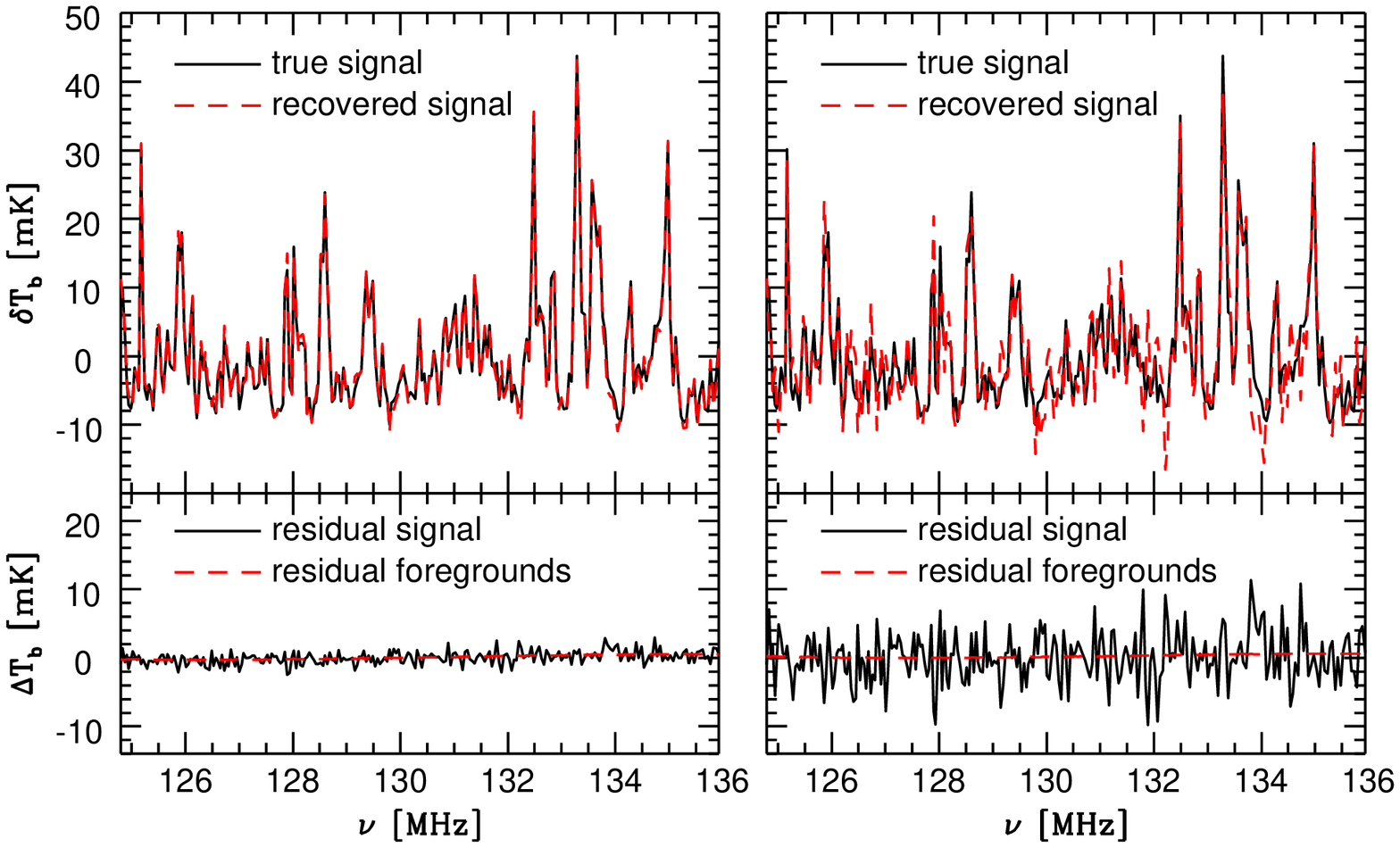}}
\caption{Same as Fig. \ref{fig:tc_avg}, but for one of the best
reconstructions, where the rms difference is $d_{\rm rms}=0.93$ for
$\sgn=1$ mK and $d_{\rm rms}=3.62$ for $\sgn=5$ mK.}
\label{fig:tc_best}
\end{figure*}

We assess the ability of the method at reconstructing the cosmological
signal using the simulation output around redshift $z=10$ in the
frequency interval $\nu=124.8-136.0$ MHz in 256 bins, and for 65536
lines-of-sight. Around this redshift the mean ionization fraction is
$\fion\approx 0.38$. The galactic and extragalactic foregrounds are
introduced as described in \S \ref{sec:foregrounds}. We also add an
uncorrelated instrumental random noise component, $\sgn$, and consider
five values from $\sgn=1$ to $250$ mK. The method is then
applied on noisy and contaminated 3D brightness temperature maps
generated from the simulation for redshift space\footnote{In redshift
space the peculiar velocity (deviation from a pure Hubble flow) is
taken into account in the relation between the emission and observed
frequency.}.

\begin{table}
\caption{The mean rms difference, $d_{\rm rms}$, between the true
cosmological signal from the simulation and the reconstructed signal
from the observed data for all lines-of-sight in redshift space. We
present the results for five values of instrumental noise where the
power index of the reconstructed signal is either $\eta=0.5$ or
$\eta=1$.}
\vspace{1mm}
\begin{center}
\begin{tabular}{cccc}
\hline
 & $\eta=0.5$ & $\eta=1.0$ \\
$\sgn$ [mK] & $\left<d_{\rm rms}\right>$ & $\left<d_{\rm rms}\right>$ \\
\hline
\hline
1 & $1.78\pm 0.57$ & $1.78\pm 0.57$ \\
5 & $4.48\pm 0.34$ & $4.27\pm 0.39$ \\
20 & $10.23\pm 0.62$ & $8.34\pm 0.77$ \\
100 & $20.17\pm 1.71$ & $13.54\pm 1.61$ \\
250 & $27.04\pm 3.17$ & $15.62\pm 2.71$ \\
\hline
\end{tabular}
\end{center}
\label{tbl:d_rms}
\end{table}

To estimate the accuracy of the signal cleaning process we calculated
the root-mean-square (rms) of the difference between the true signal
from the simulation, $\btctru$, and the reconstructed signal from the
observed data, $\btcrec$, for all the points along each line-of-sight,
$l$,
\begin{equation}
d_{\rm rms}=\sqrt{\left<\left(\btctru-\btcrec\right)^2\right>}.
\label{eq:Tcos}
\end{equation}

In Fig. \ref{fig:dsig} we present the mean $d_{\rm rms}$ for all the
lines-of-sight at redshift space as a function of the instrumental
noise. We calculated $\left<d_{\rm rms}\right>$ nine values of noise
from $\sgn=1$ to $250$ mK (full circles). We also found
the best linear fit in logarithmic scale, $\left<d_{\rm
rms}\right>=2.74\ln\sgn+0.78$. Since $\left<d_{\rm rms}\right>$ has a
linear dependence in $\ln\sgn$, we expect the results of the signal
reconstruction process in higher noise levels to be similar to the
results in the case of $\sgn=250$ mK.

In table \ref{tbl:d_rms} we present the calculation of the mean
$d_{\rm rms}$ for all the lines-of-sight, for five values of
instrumental noise $\sgn=1$, $5$, $20$, $100$ and $250$ mK, and for
$\eta=0.5$ and $1$. As expected, $\eta=1$ gives better $\left<d_{\rm
rms}\right>$ than $\eta=0.5$ and the differences become more
significant as the instrumental noise increases. In the case of
$\sgn=1$ mK, $\left<d_{\rm rms}\right>$ is higher then the noise
level. This happened because we did not take into account the errors
in the foreground fitting which has an rms value is of the order of 2
mK. The foreground error rms stays approximately 2 mK also for higher
values of $\sgn$ and therefore becomes less significant as the
instrumental noise increases.

In Fig. \ref{fig:tc_avg} we present an example of a recovered
cosmological 21-cm signal compared with the true signal along a
line-of-sight and for the noise levels: $\sgn=1$ mK (top-left panel,
$d_{\rm rms}=1.77$) and $\sgn=5$ mK (top-right panel, $d_{\rm
rms}=4.50$). The residue between the true and the recovered signals is
$|\Delta\tb|\lsim 5$ mK for $\sgn=1$ mK (bottom-left panel) and
$|\Delta\tb|\lsim 12$ mK for $\sgn=5$ mK (bottom-right panel). In
Fig. \ref{fig:tc_best} we present results for one of the best
recovered lines-of-sight, where the rms difference is $d_{\rm
rms}=0.93$ for $\sgn=1$ mK and $d_{\rm rms}=3.62$ for $\sgn=5$ mK. The
signal residue in this case is $|\Delta\tb|\lsim 2$ mK for $\sgn=1$ mK
(bottom-left panel) and $|\Delta\tb|\lsim 10$ mK for $\sgn=5$ mK
(bottom-right panel). For all cases in Figs. \ref{fig:tc_avg} and
\ref{fig:tc_best}, $\eta=0.5$. Since the cosmological signal in the
simulation box has a mean value of $\delta\tb=10.15\pm 8.61$ mK, there
was no point to present the recovered signal for a single
line-of-sight for higher values of $\sgn$.

We also compute the reconstructed signals for our range of low and
high galactic foregrounds. We calculate the mean $d_{\rm rms}$ for
these cases and found no significant differences to our ``optimal''
case.

\subsubsection{The power spectrum}
\label{sec:pk}

\begin{figure}
\centering
\mbox{\epsfig{figure=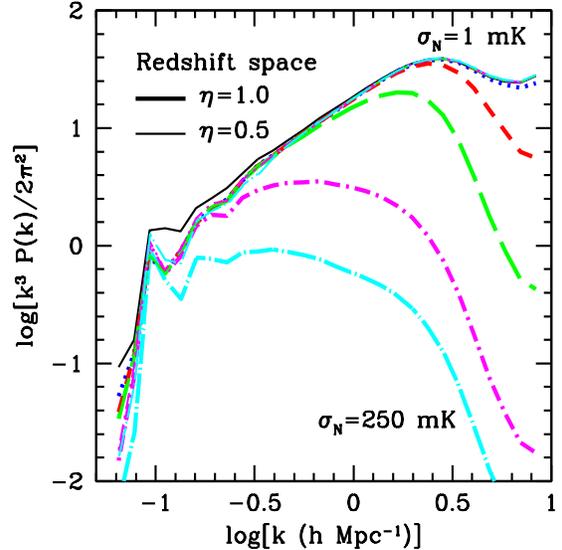,height=3.2in,width=3.2in}}
\caption{The dimensionless power spectrum, $\Delta^2(k)$, of the
brightness temperature, $\delta\tb$ in redshift space, for the
cosmological signal from the simulation (thin solid line), and the
reconstructed signal with: $\sgn=1$ mK (dotted line), $\sgn=5$ mK
(dashed line), $\sgn=20$ mK (long-dashed line), $\sgn=100$ mK
(dash-dotted line), and $\sgn=250$ mK (long-dash-dotted line), where
both $\eta=1$ (thick lines), and $\eta=0.5$ (thin lines) were used.}
\label{fig:pk256r}
\end{figure}

\begin{figure*}
\centering
\resizebox{0.96\textwidth}{!}{\includegraphics{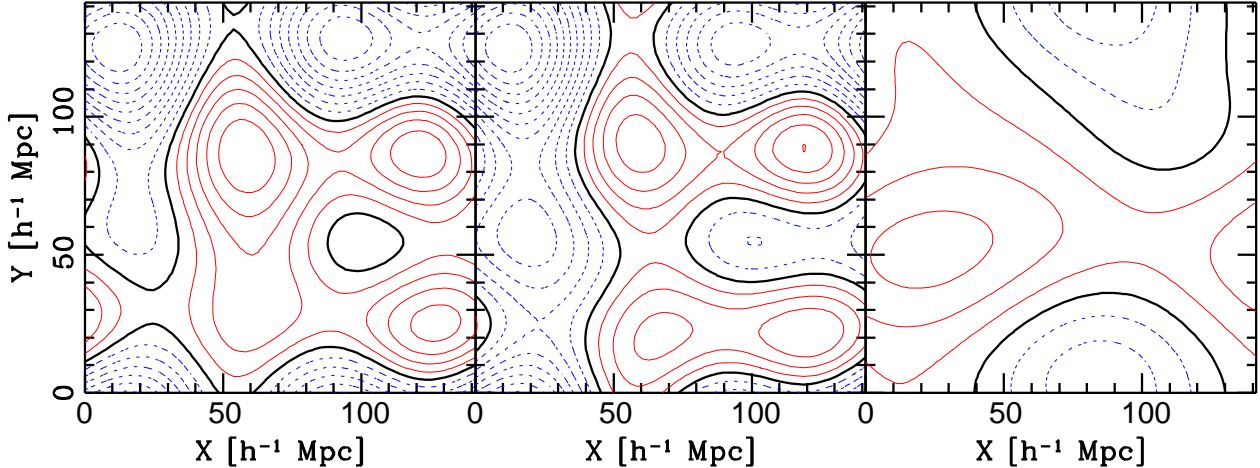}}
\caption{The left panel shows a 2D slice of the redshift space 3D map
of the true 21-cm cosmological signal from the simulation
$\Delta\tb=\delta\tb-\left<\delta\tb\right>$ at $z\approx 10$ after
applying a low-pass filter in $k$-space, where $k\le 0.1\ {\rm h\
Mpc^{-1}}$. The middle panel shows the differences between the true
foregrounds and the fitted foregrounds for the same slice and after
applying the same low-pass filter. The difference between the left and
middle panels is drawn in the right panel. The thick solid contour
indicates $\Delta\tb=0$ mK, while the thin solid and dotted contours,
respectively, represent the brightness temperatures above and below
zero. The contour spacing is 0.2 mK, the 2D slice is perpendicular to
the lines-of-sight, and $\sgn=1$ mK}
\label{fig:cthr2.3}
\end{figure*}

In Fig. \ref{fig:pk256r} we compare the dimensionless power spectrum,
$\Delta^2(k)\equiv(V/8\pi^3)4\pi k^3\Pk$, of the brightness
temperature, $\delta\tb$, between the true cosmological signal from
the simulation (thin solid line), and the reconstructed signal with:
$\sgn=1$ mK (dotted line), $\sgn=5$ mK (dashed line), $\sgn=20$ mK
(long-dashed line), $\sgn=100$ mK (dash-dotted line), and $\sgn=250$
mK (long-dash-dotted line), both for $\eta=1$ (thick lines), and
$\eta=0.5$ (thin lines). As expected from equation (\ref{eq:Pk rec}),
for $\eta=1$ the power is suppressed at small scales (large
wavenumbers $k$), while $\eta=0.5$ gives excellent fit for all
$\sgn$. At large scales, where $\log(k)\lsim-0.4$, the size of the
simulation box limited the ability to get sufficient statistics in the
foreground fitting process. Therefore, the fitted foregrounds include
some of the cosmological signal. In Fig. \ref{fig:cthr2.3} we present
a comparison of the true cosmological signal after applying a low-pass
filter in $k$-space, where $k\le 0.1\ {\rm h\ Mpc^{-1}}$, with the
differences between the true and the fitted foregrounds after applying
the same low-pass filter, left and middle panels, respectively. One
can see the similarity between the two maps which means that for large
scales a significant part of the cosmological signal is included in
the fitted foregrounds. To overcome this problem one should use a
larger simulation box.

\begin{figure*}
\centering
\resizebox{0.96\textwidth}{!}{\includegraphics{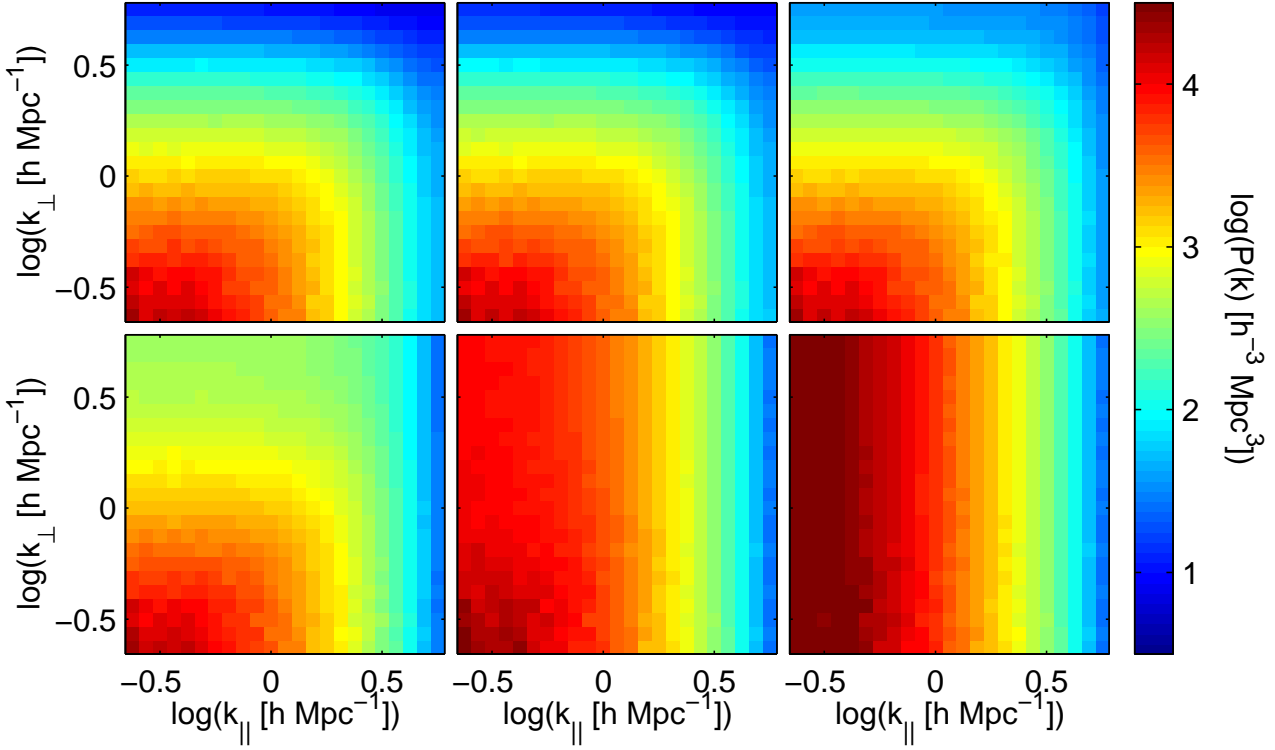}}
\caption{The power spectrum in logarithmic scale of the brightness
temperature, $\delta\tb$, in the $k_\perp-k_\parallel$ plane, where
$k_\perp$ and $k_\parallel$ are the wavenumbers perpendicular and the
parallel to the lines-of-sight, respectively. The power spectrum of
the cosmological signal from the simulation is presented in the
top-left panel, while the rest of the panels show the power spectrum
of the reconstructed signal from the observed data for five values of
$\sgn$: 1 mK (top-middle), 5 mK (top-right), 20 mK (bottom-left), 100
mK (bottom-middle), and 250 mK (bottom-right). All cases are in
redshift space, and $\eta=0.5$ was used to reconstruct the signals.}
\label{fig:cpkr}
\end{figure*}

In Fig. \ref{fig:cpkr} we present $\delta\tb$ the power spectrum in
logarithmic scale in the $k_\perp-k_\parallel$ plane, where $k_\perp$
and $k_\parallel$ are the wavenumbers perpendicular and parallel to
the lines-of-sight, respectively. The power spectrum of the
cosmological signal from the simulation in redshift space is presented
in the top-left panel, while the rest of the panels show the power
spectrum of the reconstructed signal from the observed data for five
values of $\sgn$: 1 mK (top-middle), 5 mK (top-right), 20 mK
(bottom-left), 100 mK (bottom-middle), and 250 mK (bottom-right),
where $\eta=0.5$ for all cases. One can see that as the noise level
grows the power spectrum of the reconstructed signal becomes less and
less reliable, especially for large $k_\perp$ (small scales), where
the high instrumental noise eliminates the differences between large
and small $k_\perp$.

\begin{figure}
\centering
\mbox{\epsfig{figure=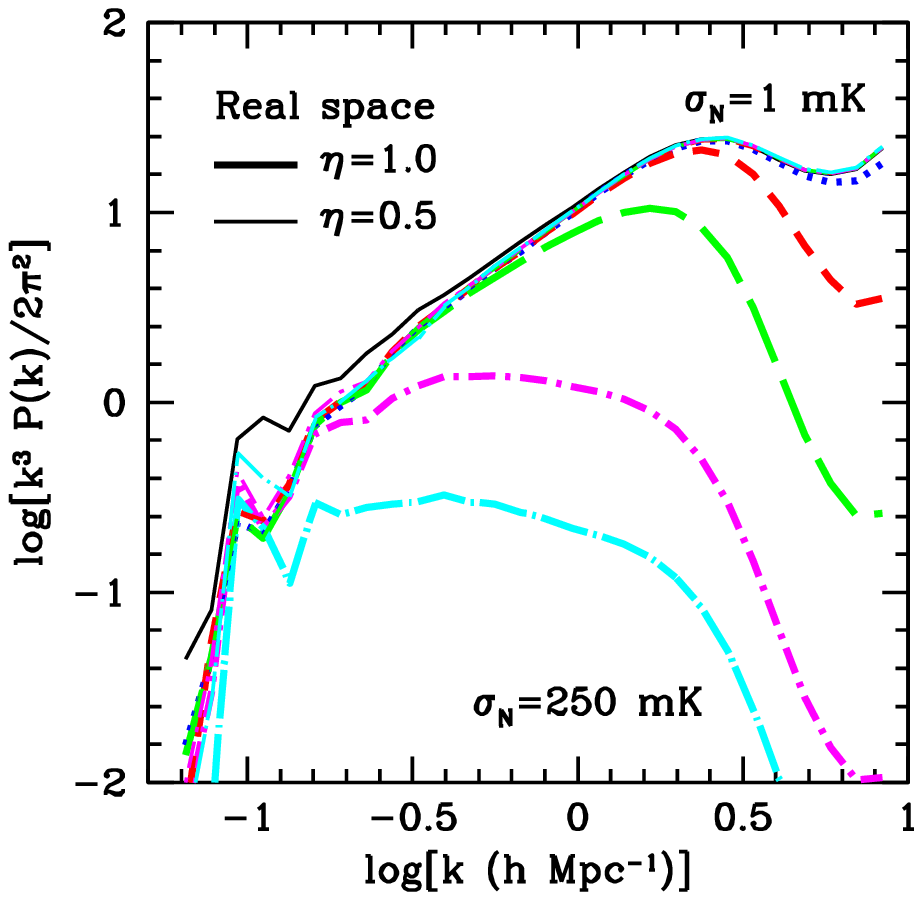,height=3.2in,width=3.2in}}
\caption{The same as Fig. \ref{fig:pk256r} but in real space.}
\label{fig:pk256d}
\end{figure}

\begin{figure*}
\centering
\resizebox{0.96\textwidth}{!}{\includegraphics{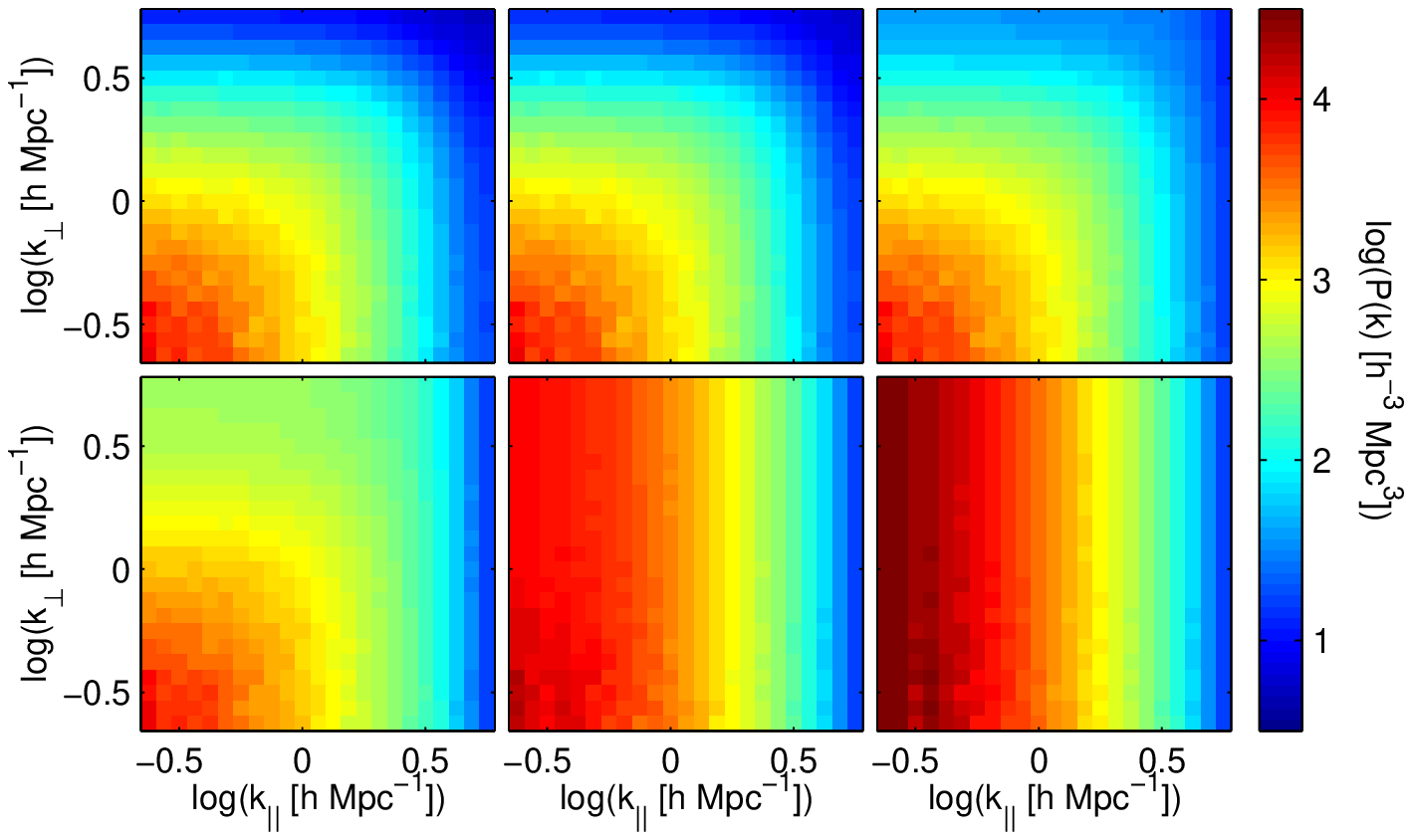}}
\caption{The same as Fig. \ref{fig:cpkr} but in real space.}
\label{fig:cpkd}
\end{figure*}

Nusser 2005b suggested to apply the Alcock-Paczy\'nski test on the
observed maps of the cosmological signal (cf. Pandey \& Bhardwaj 2005;
Barkana \& Loeb 2005). To do so one must compare the power spectrum in
redshift space with power spectrum in real space. Figs. \ref{fig:pk256d}
and \ref{fig:cpkd} are the same as Figs. \ref{fig:pk256r} and
\ref{fig:cpkr}, respectively, but for real space. The dimensionless
power spectrum, $\Delta^2(k)$, is slightly lower in real space, and
the two-dimensional (2D) power spectrum is symmetric in real space
and asymmetric in redshift space. We see that a successful application
of this test requires noise levels significantly lower than 100 mK.

\begin{figure*}
\centering
\resizebox{0.96\textwidth}{!}{\includegraphics{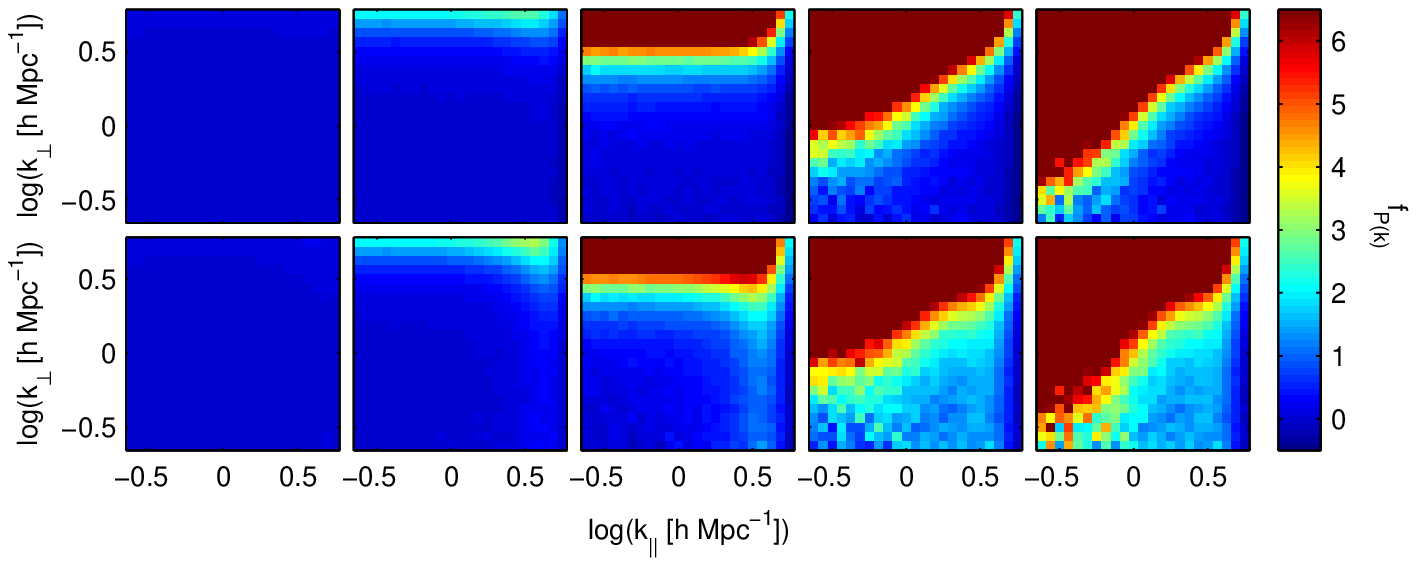}}
\caption{The ratio, $f_{_{\rm P(k)}}\equiv P_{\rm rec}(k)/P_{\rm
sig}(k)-1$, between the recovered and true signal power spectrum in
redshift space (top panels) and real space (bottom panels). The
results for five values of $\sgn$: 1, 5, 20, 100, and 250 mK are
presented in five panels from left to right, respectively. For all
cases $\eta=0.5$.}
\label{fig:cdpk2}
\end{figure*}

\begin{table*}
\caption{The mean differences (in percentages) between the true
cosmological signal from the simulation and the reconstructed signal
from the observed data for five values of $\sgn$ and three different
maximal $\log(k)$, both in redshift space and real space. The Wiener
filter parameter is $\eta=0.5$ for all cases.}
\vspace{1mm}
\begin{center}
\begin{tabular}{ccccccc}
\hline
 & \multicolumn{3}{c}{Redshift space} & \multicolumn{3}{c}{Real space} \\
 & \multicolumn{6}{c}{$\log(k)$ [h Mpc$^{-1}$]} \\
$\sgn$ [mK] & $\lsim 0.03$  & $\lsim 0.21$ & $\lsim 0.51$ & $\lsim 0.03$ & $\lsim 0.21$ & $\lsim 0.51$ \\
\hline
\hline
1 & $-0.4\pm0.9$ & $-0.3\pm0.8$ & $0.0\pm0.8$ & $-0.1\pm1.2$ & $-0.1\pm1.0$ & $0.3\pm1.0$ \\
5 & $0.2\pm1.4$ & $0.8\pm1.7$ & $5.0\pm8.7$ & $0.8\pm1.9$ & $1.8\pm2.2$ & $8.7\pm12.4$ \\
20 & $9.4\pm8.7$ & $17.6\pm18.7$ & $71.7\pm114.5$ & $17.1\pm11.6$ & $29.3\pm21.3$ & $100.7\pm124.7$ \\
100 & $180.5\pm159.0$ & $291.1\pm345.0$ & $912.3\pm1797.4$ & $254.2\pm143.1$ & $350.4\pm300.0$ & $885.6\pm1568.6$ \\
250 & $574.9\pm640.3$ & $881.2\pm1336.1$ & $2627.0\pm6452.1$ & $620.4\pm500.5$ & $821.7\pm1018.4$ & $2060.9\pm4841.7$ \\
\hline
\end{tabular}
\end{center}
\label{tbl:fpk}
\end{table*}

We also present (Fig. \ref{fig:cdpk2}) the ratio, $f_{_{\rm
P(k)}}\equiv P_{\rm rec}(k)/P_{\rm sig}(k)-1$, in the
$k_\perp-k_\parallel$ plane, between the power spectrum of the true
cosmological signal, $P_{\rm sig}(k)$, and the power spectrum of the
reconstructed signal, $P_{\rm rec}(k)$, where the top and bottom
panels are for redshift and real space, respectively. We present the
results for the same five values of $\sgn$: 1, 5, 20, 100, and 250 mK
in five panels from left to right, respectively. For low instrumental
noise of $\sgn=1$ mK the results are reliable with no significant
differences on average between the true and reconstructed signals and
a variance of $\sim 1\%$ for $\log(k)\lsim0.5$ [h Mpc$^{-1}$], both in
real and redshift space. The results for $\sgn=5$ mK are still
reasonably good with differences between the true and reconstructed
signals of $5.0\pm8.7\%$ in redshift space and $8.7\pm12.4\%$ in real
space, both for $\log(k)\lsim0.5$ [h Mpc$^{-1}$]. As the instrumental
noise increases the differences between the true and reconstructed
signals increases and the results becomes less and less reliable (for
more details see table \ref{tbl:fpk}).

\subsubsection{The Minkowski functionals}
\label{sec:mfs}

\begin{figure}
\centering
\mbox{\epsfig{figure=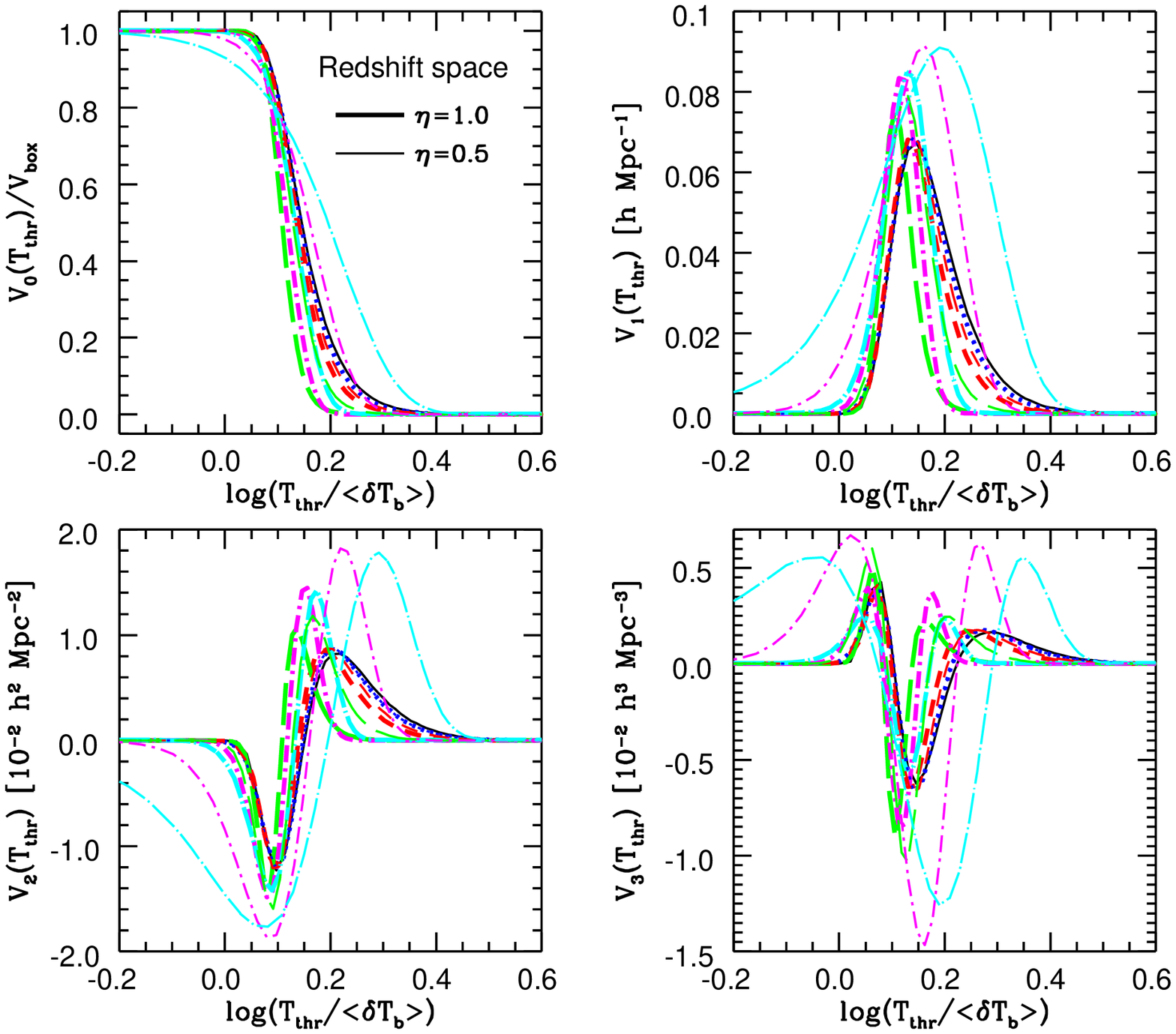,height=2.9in,width=3.2in}}
\caption{The Minkowski functionals in redshift space of the true
cosmological signal brightness temperature (thin solid line), and the
reconstructed signal with: $\sgn=1$ mK (dotted line), $\sgn=5$ mK
(dashed line), $\sgn=20$ mK (long-dashed line), $\sgn=100$ mK
(dash-dotted line), and $\sgn=250$ mK (long-dash-dotted line), where
both $\eta=1$ (thick lines), and $\eta=0.5$ (thin lines) were used.}
\label{fig:mf256r}
\end{figure}

\begin{figure}
\centering
\mbox{\epsfig{figure=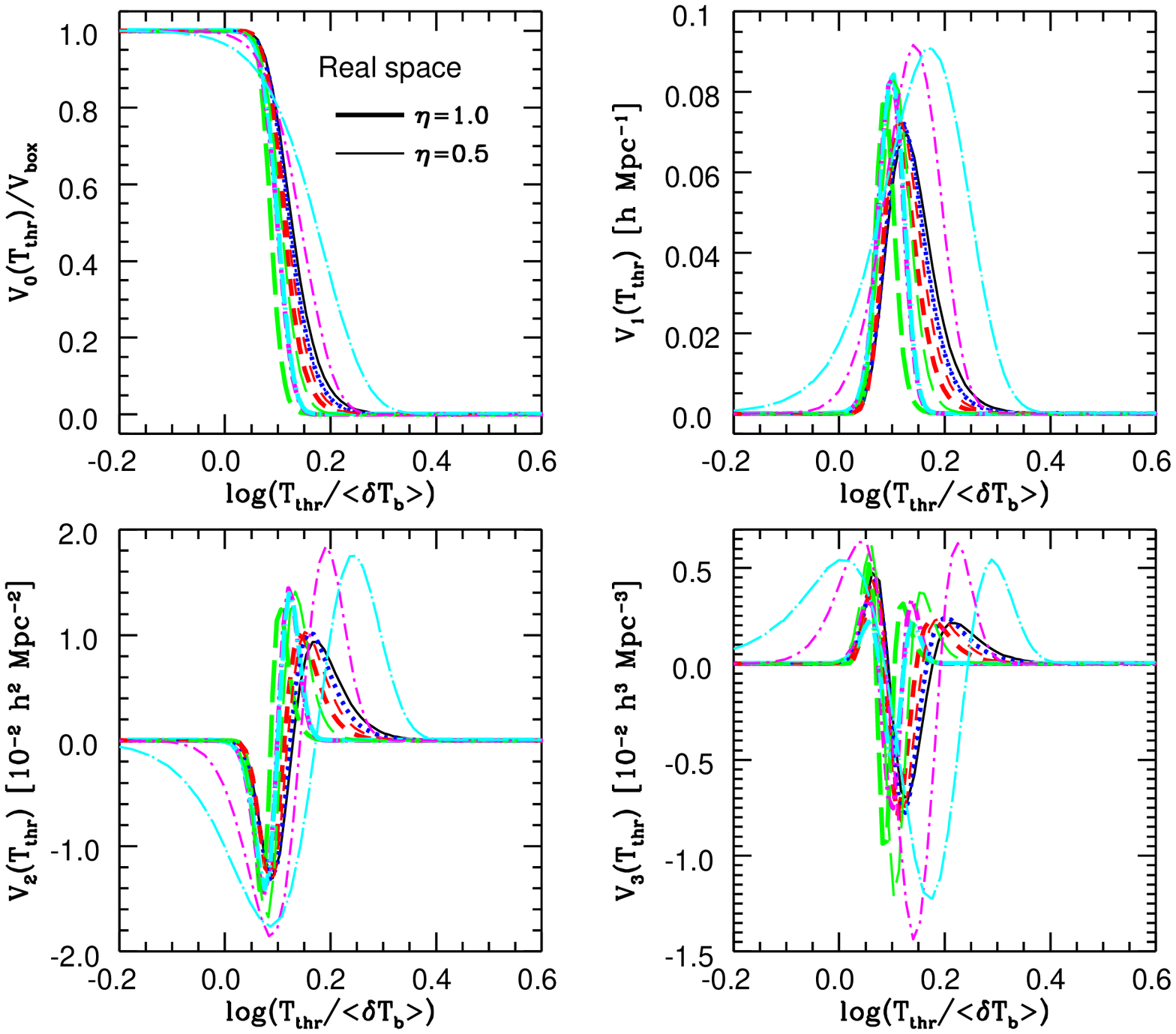,height=2.9in,width=3.2in}}
\caption{The same as Fig. \ref{fig:mf256r} but in real space.}
\label{fig:mf256d}
\end{figure}

The morphology of the total gas and the neutral hydrogen distributions
during the cosmological epoch of reionization can be quantified with
Minkowski functionals (MFs) of isodensity surfaces. Gleser \etal
(2006) suggested the MFs as a means of investigating the reionization
epoch. In Figs. \ref{fig:mf256r} and \ref{fig:mf256d} we present the
MFs from the 21-cm cosmological signal in redshift and real space,
respectively, where $V_0(\nu)$ is the volume of regions with
brightness temperature above the threshold $T_{\rm thr}$, $V_1(\nu)$
is a measure of the surface area of the boundary, $V_2(\nu)$ is the
mean curvature over the surface, and $V_3(\nu)$ is the Euler
characteristic, $\chi$. We compare the MFs of the true cosmological
signal brightness temperature (thin solid line), and the reconstructed
signal with: $\sgn=1$ mK (dotted line), $\sgn=5$ mK (dashed line),
$\sgn=20$ mK (long-dashed line), $\sgn=100$ mK (dash-dotted line), and
$\sgn=250$ mK (long-dash-dotted line), both for $\eta=1$ (thick
lines), and $\eta=0.5$ (thin lines). Since the MFs are sensitive to
the residual noise and foreground contamination, we smooth the 21-cm
maps with a Gaussian filter of width $0.55$ h$^{-1}$Mpc before
applying the MFs computing code to them. 

A comparison of the $V_0$ curves in the figures shows that the
reconstructed maps have fewer regions with high brightness temperature
than the true cosmological signal. This effect increases with the
noise rms $\sgn$. Both in real and redshift space, $\eta=0.5$ perform
better than $\eta=1$ at recovering the true MFs. In any case, the
recovery of the MFs is reasonable for $\sgn=1$ and $5$ mK, suggesting
that in the future, when the noise level will be reduced to
$\sgn\approx5$ mK, the MFs could be used to distinguish between
different stages of ionization and between different reionization
scenarios as advocated by Gleser et al. (2006).


\section{Discussion}
\label{sec:discussion}

Maps of 21-cm radiation from diffuse \hi\ before and during the epoch
of reionization are potentially our only direct probe of the Universe
in the dark ages. From this point of view the efforts invested in
modeling this radiation and in attempts to observe it are worthwhile.
Redshifted 21-cm maps are likely to constrain the early stages of star
and galaxy formation and also the state of the diffuse \hi\ before any
light emitting object had formed. Decoupling of the \hi\ spin
temperature, $\ts$, from the CMB temperature, $\tcmb$, is necessary
for observing redshifted 21-cm (either in absorption or in emission).
In the absence of any heating mechanisms, the kinetic gas temperature,
$\tk$, drops below $\tcmb$ at redshifts $z\lsim 100$. For $z\gsim 20$
the Universe is dense enough so that atom-atom collisions are still
efficient at bringing $\ts$ down towards $\tk$. However, even in the
absence of luminous structure, the gas temperature could be raised
above the CMB if the dark matter could decay or annihilate
(e.g. Ripamonti, Mapelli \& Ferrara 2007). For example, neutralinos,
if they are the dark matter particles, produce sufficient
electron-positron $e^-e^+$pairs by self-annihilations (Myers \& Nusser
2007). High energy photons generated by inverse Compton scattering of
CMB photons, could then photo-heat the gas to kinetic temperatures
above $\tcmb$. In this case, the gas is never seen in
absorption. Therefore, 21-cm maps could in principle also constrain
the nature of the dark matter particle.

The 21-cm cosmological signal will suffer from several sources of
contamination. In the first part of this work we model the
contamination which is produced by foreground radiation emitted from
galactic and extragalactic sources, and produced 3D contamination
maps. The foregrounds model includes synchrotron and free-free emission
from our Galaxy and seven extragalactic foregrounds: radio point
sources which relate to AGN activity, radio haloes and radio relics
from massive dark matter haloes with mass of $M_{\rm halo}\gsim
5\times 10^{14}\msun$, synchrotron and free-free emission from star
forming galaxies, and free-free emission from ionized hydrogen dark
matter haloes and diffuse IGM. The most significant foreground is the
galactic synchrotron which cause $\sim$90\% of the contamination on
average. The galactic free-free contributes another 1-2\% to the total
contamination. The AGN radio point sources produce $\sim$10\% of the
total contamination on average, but can reach $\sim$25\% at high
galactic latitudes, where the minimum brightness temperature of the
diffuse galactic emission drops to $\sim$200 K. Radio haloes and radio
relics are also significant foregrounds, but since they appear only in
rich galaxy clusters, they are rare and should appear only as spots in
individual lines-of-sight. The remaining extragalactic foregrounds are
less significant and contribute less the 1\% to the total
contamination.

In the second part we add the 3D contamination maps and an
instrumental random noise to the maps of the cosmological 21-cm
signal, and apply a reconstruction method based on a Bayesian
statistical approach. We take advantage of the foregrounds' smooth
dependence on frequency in contrast to the ``noisy'' signal. Then we
derive an appropriate Wiener filter to extract the cosmological signal
from the residual brightness temperature obtained after subtracting
the approximated foregrounds from the observed data. We also introduce
a power index $\eta$ which determines the strength of the Wiener
filter. The Wiener filter obtained directly from Bayesian statistics
has $\eta=1$, but using $\eta=0.5$ eliminates the power suppression at
small scales. For low instrumental noise we were able to get
reasonable reconstruction of the cosmological signal for each
line-of-sight, where $d_{\rm rms}\approx 1.7\pm 0.6$ for $\sgn=1$ mK,
and $d_{\rm rms}\approx 4.2\pm 0.4$ for $\sgn=5$ mK. We checked the
reconstruction of the statistical measurements of the power spectrum
and the MFs. The 1D power spectrum was nicely reconstructed for all
values of the instrumental noise up to 250 mK, while the 2D power
spectrum and the MFs were reconstructed reasonably well only for
noise levels significantly lower than 100 mK.


\section*{Acknowledgments}

LG and AN acknowledge the support of the Asher Space Research
Institute and the German Israeli Foundation for the Development of
Research. AJB acknowledges support from the Gordon \& Betty Moore
Foundation.




\appendix


\section{The method of signal reconstruction}
\label{apx:signalrec}

The maximum a-posteriori probability (MAP) formalism have been used to
reconstruct the large-scale structure of the Universe (e.g. Rybicki \&
Press 1992; Zaroubi \etal 1995; Fisher \etal 1995). Based on this
formalism, we develop a method to reconstruct the weak 21-cm
cosmological signal from the contaminated noisy data. The method
relies on the smoothness of the contaminating radiation along the
frequency axis and an assumed prior for the correlation properties of
the cosmological signal.


\subsection{The conditional probability}
\label{apx:probability}

Let $P(\bs|\bd)$ be the conditional probability distribution function
to get the signal vector $\bs$ given the data vector $\bd$. The MAP
method estimates $\bs$ by maximization of $P(\bs|\bd)$. According to
Bayes theorem
\begin{equation}
P(\bs|\bd)=\frac{P(\bd|\bs)P(\bs)}{P(\bd)}.
\label{eq:bayes}
\end{equation}
Since the maximization is on $\bs$, the denominator $P(\bd)$ can be
ignored. In addition to the true signal, the data $\bd$ contains
contamination from different foreground sources and a random
(instrumental) noise. Here $\bc$ represents the fit to the total
contamination. The $i$th element of the contamination along a
line-of-sight $l$ is
\begin{equation}
c_{_{l,i}}=\exp\left(\sum_{m=0}^{n_{_m}}b_{_{l,m}}x_{_{l,i}}^m\right)
\label{eq:cont}
\end{equation}
where we use a polynomial approximation in
$x_{_{l,i}}=\ln(\nu_{_{l,i}}/\nu_{_\ast})$ of order $n_{_m}$,
$\nu_{_{l,i}}$ is the frequency at position $i$ along the
line-of-sight $l$, $\nu_{_\ast}$ is a constant frequency, and
$b_{_{l,m}}$ are the polynomial coefficients.\footnote{We use two
indices $l$ and $i$ to determine the position of each element in the
vectors $\bc$ and $\bx$. One can replace $l$ and $i$ with a single
index $j$ where $j=\ni(l-1)+i$ and $\ni$ is the number of frequencies
along a single line-of-sight.} Usually, the frequency
$\nu_{_{l,i}}\equiv\nu_{_i}$ is the same for all $l$'s, and therefore,
the vector $\bx$ contains $\nl$ times the vector
$x_{_i}=\ln(\nu_{_i}/\nu_{_\ast})$, where $\nl$ is the number of
lines-of-sight. We determine the residue vector $\by=\bd-\bc$ as the
residual data $\bd$ after subtraction of the approximated
contaminations $\bc$. So ideally, $\by$ contains only the true signal
and the random noise.

Assuming that the signal is approximately a Gaussian random field with
an auto-correlation matrix $\bS$, the probability function of $\bs$
is
\begin{equation}
P(\bs)=\exp\left(-\frac{1}{2}\bs^{\rm T}\bSn\bs\right).
\label{eq:Ps}
\end{equation}
Further, assuming a Gaussian noise matrix $\bN$, the probability for
$\bd$ giving $\bs$ is
\begin{equation}
P(\bd|\bs)=\exp\left[-\frac{1}{2}(\by-\bs)^{\rm T}\bNn(\by-\bs)\right]\; .
\label{eq:Pds}
\end{equation}
In the following we assume $\bN=\bI\sgn$, where $\sgn$ is the noise
rms and $\bI$ is the identity matrix.

Using the above probability functions (equations (\ref{eq:Ps}) \&
(\ref{eq:Pds})), the probability $P(\bs|\bd)$ is proportional to
\begin{align}
\label{eq:Psd1}
P(\bs|\bd) & \propto P(\bd|\bs)P(\bs)\nonumber\\
 & =\exp\left\{-\frac{1}{2}\left[(\by-\bs)^{\rm T}\bNn(\by-\bs)+\bs^{\rm T}\bSn\bs\right]\right\}\nonumber\\
 & = \exp\left\{-\frac{1}{2}\left[\by^{\rm T}\bNn\by+\bs^{\rm T}(\bSNn)\bs\right.\right.\nonumber\\
 & \ \ \ \ \ \ \ \ \ \ \ \left.\left.-\by^{\rm T}\bNn\bs-\bs^{\rm T}\bNn\by\right]\right\},\nonumber\\
 & = \exp\left\{-\frac{1}{2}\left[\by^{\rm T}(\bSN)^{-1}\by+\bs^{\rm T}(\bSNn)\bs\right.\right.\nonumber\\
 & \ \ \ \ \ \ \ \ \ \ \ +\by^{\rm T}\bNn\bS(\bSN)^{-1}\by\nonumber\\
 & \ \ \ \ \ \ \ \ \ \ \ -\by^{\rm T}\bS(\bSN)^{-1}(\bSNn)\bs\nonumber\\
 & \ \ \ \ \ \ \ \ \ \ \ \left.\left.-\bs^{\rm T}(\bSNn)\bS(\bSN)^{-1}\by\right]\right\},
\end{align}
where we use the matrix identity
\begin{align}
\label{eq:mtxidf2}
\bNn &= \bNn(\bSN)(\bSN)^{-1}=(\bNn\bS+\bI)(\bSN)^{-1}\nonumber\\
 &= \bNn\bS(\bSN)^{-1}+(\bSN)^{-1}
\end{align}
for the first term in the rhs of the equation, and 
\begin{equation}
\label{eq:mtxidf1}
(\bSNn)^{-1}=\bS(\bSN)^{-1}\bN=\bN(\bSN)^{-1}\bS
\end{equation}
for the third and the fourth terms in the rhs of the equation. Next we
again use the identity in equation (\ref{eq:mtxidf1}) for the third
term in the rhs of equation (\ref{eq:Psd1}), and get
\begin{align}
\label{eq:Psd2}
P(\bs|\bd) & \propto\exp\left\{-\frac{1}{2}\left[\by^{\rm T}(\bSN)^{-1}\by+\bs^{\rm T}(\bSNn)\bs\right.\right.\nonumber\\
 & \ \ \ \ \ \ \ \ \ \ \ +\by^{\rm T}\bS(\bSN)^{-1}(\bSNn)\bS(\bSN)^{-1}\by\nonumber\\
 & \ \ \ \ \ \ \ \ \ \ \ -\by^{\rm T}\bS(\bSN)^{-1}(\bSNn)\bs\nonumber\\
 & \ \ \ \ \ \ \ \ \ \ \ \left.\left.-\bs^{\rm T}(\bSNn)\bS(\bSN)^{-1}\by\right]\right\}\nonumber\\
 & =\exp\left\{-\frac{1}{2}\left[\by^{\rm T}(\bSN)^{-1}\by+\bs^{\rm T}(\bSNn)\bs\right.\right.\nonumber\\
 & \ \ \ \ \ \ \ \ \ \ \ +\left(\bS(\bSN)^{-1}\by\right)^{\rm T}(\bSNn)\left(\bS(\bSN)^{-1}\by\right)\nonumber\\
 & \ \ \ \ \ \ \ \ \ \ \ -\left(\bS(\bSN)^{-1}\by\right)^{\rm T}(\bSNn)\bs\nonumber\\
 & \ \ \ \ \ \ \ \ \ \ \ \left.\left.-\bs^{\rm T}(\bSNn)\left(\bS(\bSN)^{-1}\by\right)\right]\right\}\nonumber\\
 & =\exp\left\{-\frac{1}{2}\left[\by^{\rm T}(\bSN)^{-1}\by+\left(\bs-\bS(\bSN)^{-1}\by\right)^{\rm T}\right.\right.\nonumber\\
 & \ \ \ \ \ \ \ \ \ \ \ \left.\left.\times(\bSNn)\left(\bs-\bS(\bSN)^{-1}\by\right)\right]\right\},
\end{align}


\subsection{Fitting the contamination}
\label{apx:contamination}

The fitting process of the total contamination, $\bc$, depends on two
assumptions: ($i$) the signal is much weaker then the total
contamination, and ($ii$) the contamination has a smooth frequency
spectrum while the signal is rapidly changing in frequency. Once these
assumptions are valid, we can switch between the signal and the
contamination, and treat the contamination as the desirable signal and
the signal as an additional random noise. Now, the polynomial
coefficients, $b_{_{l,m}}$, can be calculated by minimization of the
first term of the exponent at the rhs of the conditional probability
distribution function (equation (\ref{eq:Psd2}))
\begin{subequations}
\begin{equation}
q=\frac{1}{2}\sum_{l=1}^\nl\sum_{i=1}^\ni\sum_{l'=1}^\nl\sum_{i'=1}^\ni(d_{_{l,i}}\!\!-\!\!c_{_{l,i}})(\bSN)_{l,i,l',i'}^{-1}(d_{_{l',i'}}\!\!-\!\!c_{_{l',i'}})\ ,
\label{eq:ql}
\end{equation}
\begin{equation}
\frac{\partial q}{\partial b_{_{l,m}}}=-\sum_{i=1}^\ni\sum_{l'=1}^\nl\sum_{i'=1}^\ni x_{_{l,i}}^mc_{_{l,i}}(\bSN)_{l,i,l',i'}^{-1}(d_{_{l',i'}}\!\!-\!\!c_{_{l',i'}})\ ,
\label{eq:dql}
\end{equation}
\begin{align}
\label{eq:d2ql}
\frac{\partial^2 q}{\partial b_{_{l,m}}\partial b_{_{l',n}}} =& \ \sum_{i=1}^\ni\sum_{i'=1}^\ni\left[x_{_{l,i}}^mc_{_{l,i}}(\bSN)_{l,i,l',i'}^{-1}x_{_{{l',i'}}}^nc_{_{l',i'}}\right.\nonumber\\
 &\ \left.-\delta_{_{l-l'}}x_{_{l,i}}^{m+n}c_{_{l,i}}(\bSN)_{l,i,l',i'}^{-1}(d_{_{l',i'}}\!\!-\!\!c_{_{l',i'}})\right]
\end{align}
\end{subequations}
where $i$ and $i'$ are the frequency position indices, from 1 to
$\ni$, along the lines-of-sight $l$ and $l'$, respectively. The number
of lines-of-sight is $\nl$, and $\delta_{_{l-l'}}$ is the Kronecker
delta which equal one for $l=l'$ and zero otherwise. Equations
(\ref{eq:dql}) and (\ref{eq:d2ql}) are the elements of the gradient
and Hessian, respectively, where the are $n_{_m}+1$ coefficients,
$b_{_{l,m}}$, for each line-of-sight.

The minimization can be calculated also in Fourier-space
\begin{subequations}
\begin{equation}
{\mathcal Q}=\frac{1}{2}\sum_k\frac{{\mathcal Y}^2_k}{\Pk+\sgn^2}=\frac{1}{2}\sum_k\frac{\left({\mathcal D}_k-{\mathcal C}_k\right)^2}{\Pk+\sgn^2},
\label{eq:Ql}
\end{equation}
\begin{equation}
\frac{\partial {\mathcal Q}}{\partial b_{_{l,m}}}=-\sum_k\frac{{\mathcal F}_k\left(\bz_{_m}\right)\left({\mathcal D}_k-{\mathcal C}_k\right)}{\Pk+\sgn^2},
\label{eq:dQl}
\end{equation}
\begin{align}
\label{eq:d2Ql}
\frac{\partial^2 {\mathcal Q}}{\partial b_{_{l,m}}\partial b_{_{l',n}}} =& \ \sum_k\left[\frac{{\mathcal F}_k\left(\bz_{_m}\right){\mathcal F}_k\left(\bz_{_n}\right)}{\Pk+\sgn^2}\right.\nonumber\\
 & \left.- \delta_{_{l-l'}}\frac{{\mathcal F}_k\left(\bz_{_{m+n}}\right)\left({\mathcal D}_k-{\mathcal C}_k\right)}{\Pk+\sgn^2}\right],
\end{align}
\end{subequations}
where ${\mathcal Y}_k$, ${\mathcal D}_k$, and ${\mathcal C}_k$ are the
$k$ elements of the Fourier transform of $\by$, $\bd$, and $\bc$,
respectively, $\Pk$ is the field power spectrum, and ${\mathcal
F}_k\left(\bz_{_m}\right)$ is the $k$ element of the Fourier transform
of $\bz_{_m}$, where the $l'',i$ element of $\bz_{_m}$ is equal to
$x_{_{l'',i}}^mc_{_{l'',i}}$ for $l''=l$ and zero otherwise.


\subsection{The reconstructed signal}
\label{apx:recsignal}

To get the optimal reconstructed signal, $\bs$, one should minimize
the conditional probability distribution function $\partial
P(\bs|\by)/\partial\bs=0$,
\begin{equation}
\bs=\bS(\bSN)^{-1}\by,
\label{eq:bs}
\end{equation}
or in Fourier-space, $\partial P/\partial{\mathcal S}=0$,
\begin{equation}
{\mathcal S}_{k}=\frac{\Pk}{\Pk+\sgn^2}{\mathcal Y}_{k},
\label{eq:bSk}
\end{equation}
where ${\mathcal S}_{k}$ is the $k$ element of the Fourier transform
of $\bs$, and $\Pk/(\Pk+\sgn^2)$ is the Wiener filter.


\end{document}